\documentclass[10pt]{iopart}

\usepackage[utf8]{inputenc}
\usepackage{bm}
\usepackage{footnote}
\usepackage{threeparttable}
\usepackage{float}
\usepackage{amssymb}
\usepackage{amsmath}
\usepackage{physics}
\usepackage{xcolor}
\usepackage{graphicx}
\usepackage{cite}
\usepackage{siunitx}

\usepackage{hyperref}
\usepackage{todonotes}
\usepackage{upgreek}

\usepackage{xspace}

\newcounter{bla}

\usepackage{color} 

\begin{document}

\title[FEDM code]{Introduction and verification of FEDM, an open-source FEniCS-based discharge modelling code}

\author{Aleksandar P. Jovanovi\'c, Detlef Loffhagen and Markus M. Becker}

\address{Leibniz Institute for Plasma Science and Technology (INP), Felix-Hausdorff-Str. 2, 17489 Greifswald, Germany}
\ead{aleksandar.jovanovic@inp-greifswald.de}

\begin{abstract}
This paper introduces the FEDM (Finite Element Discharge Modelling) code, which was developed using the open-source computing platform FEniCS (\url{https://fenicsproject.org}).
Building on FEniCS, the FEDM code utilises the finite element method to solve partial differential equations.
It extends FEniCS with features that allow the automated implementation and numerical solution of fully-coupled fluid-Poisson models including  an arbitrary number of particle balance equations.
The code is verified using the method of exact solutions and benchmarking.
The physically based examples of a time-of-flight experiment, a positive streamer discharge in atmospheric-pressure air and a low-pressure glow discharge in argon are used as rigorous test cases for the developed modelling code and to illustrate its capabilities.
The performance of the code is compared to the commercial software package COMSOL Multiphysics\textsuperscript{\textregistered} and a comparable parallel speed-up is obtained.
It is shown that the iterative solver implemented by FEDM performs particularly well on high-performance compute clusters.
\end{abstract}
\vspace{2pc}

\ioptwocol

\section{Introduction}
\label{intro}

Electric discharges in gases producing non-thermal (cold) plasmas are used in a large variety of technical processes and applications ranging from surface processing, gas conversion and agriculture to plasma medicine~\cite{Massines-2012-ID3544, Cvelbar-2018-ID5129, Brandenburg-2018-ID5134,vonWoedtke-2013-ID3925, Weltmann-2017-ID4788, Bekeschus-2018-ID5112}.
Applications of cold physical plasmas have recently received a new impetus due to the possibility to use them to support the healthcare sector during viral pandemics such as the COVID-19 crisis~\cite{Bekeschus-2020-ID5451,Bisag-2020-ID5452}.
The computational analysis of electric discharges by means of so-called fluid-Poisson models has a long tradition~\cite{Graves-1970-ID516,Barnes-1987-ID544,Boeuf-1987-ID545,Lister-1992-ID3053,Boeuf-1995-ID3041,vanDijk-2009-ID2560,Lowke-2013-ID3112,Alves-2012-ID3406,Alves-2018-ID5094}. Such models are often applied to obtain a deeper understanding of physical processes, to support experimental investigations and to optimise processes and devices. Compared to kinetic (particle) methods~\cite{Donko-2006-ID2349,Donko-2011-ID2668, Petrovic-2017-ID4095,Loffhagen-2009-ID2590}, the advantage of fluid models is their computational efficiency, wide applicability, and ability to incorporate various physical and chemical processes.

Fluid-Poisson models for non-thermal plasmas usually consist of a set of balance equations for the particle number densities of all relevant plasma species, the Poisson equation for the electric potential, and an electron energy balance equation for determining the mean electron energy.  The latter is required in the framework of the local mean energy approximation (LMEA), which has been established as an alternative to the local field approximation (LFA) for the determination of electron transport and rate coefficients~\cite{Park-1990-ID656,Hagelaar-2005-ID2276,Grubert-2009-ID2551}. This basic set of equations may be accompanied by further relations describing relevant physical processes and must be closed by appropriate initial conditions and boundary conditions describing the plasma-surface interaction~\cite{Hagelaar-2000-ID1480,Lafleur-2019-ID5334,Bonitz-2019-ID5336}. The coupled set of partial differential equations can be solved using different discretisation methods, such as the finite difference, finite element, or finite volume method. In plasma modelling, the finite difference method has frequently been applied to simpler problems~\cite{Barnes-1988-ID582,Boeuf-1995-ID3041,Becker-2013-ID3200}, while finite volume and finite element methods have been mostly used for more complex geometries~\cite{Georghiou-2005-ID2241,Sakiyama-2010-ID2625,Li-2012-ID2887,Duarte-2015-ID3541,Trelles-2018-ID5275}.

While the basic equations included in fluid models and the applied discretisation methods have not changed much during the last decades, the progress of available hardware and software has opened up new modelling possibilities, which can only be explored by the application of modern computing and parallelisation frameworks. Currently, there are multiple open-source platforms available, such as MOOSE (Multiphysics Object Oriented Simulation Environment)~\cite{permann2019moose}, MFEM (Modular Finite Element Method library)~\cite{mfem-library}, Afivo (Adaptive Finite Volume Octree)~\cite{Teunissen-2018-ID5413}, FEniCS~\cite{FEniCS,FEniCS-book}, 
or OpenFOAM (Open-source Field Operation And Manipulation)~\cite{OpenFOAM}, which can potentially be used to create fluid-Poisson plasma modelling codes.  Nevertheless, the use of available open-source libraries for modelling of non-thermal plasmas is not very common, and only few  reports are available in the literature~\cite{Lindsay-2016-ID5394, Hromadka-2016-ID5393, Abdollahzadeh-2016-ID5410, Teunissen-2017-ID4242, Verma-2021-ID5677, Semenov-2022-ID6018}. The Zapdos application has been developed on the basis of the MOOSE framework and applied for fluid modelling of DC discharges~\cite{Lindsay-2016-ID5394}. The FEniCS platform has been used in Ref.~\cite{Hromadka-2016-ID5393} for investigation of Langmuir probe characteristics in a low-pressure argon discharge.
In Ref.~\cite{Abdollahzadeh-2016-ID5410}, a model for plasma actuators describing plasma and flow has been developed using OpenFOAM and Afivo has been used for modelling of streamer discharges~\cite{Teunissen-2017-ID4242,Bagheri-2018-ID5240}. Recently, SOMAFOAM, a finite volume framework for low temperature plasma modelling based on OpenFOAM, has been introduced in Ref.~\cite{Verma-2021-ID5677}. Furthermore, a spectral-element-based code has been developed in Ref.~\cite{Semenov-2022-ID6018} and used for the modelling of streamers.

The present manuscript presents the time-dependent and spatially two-di\-men\-sio\-nal \textbf{F}inite \textbf{E}lement 
\textbf{D}ischarge \textbf{M}odelling (FEDM) code. The code was developed on the basis of the FEniCS open-source computing platform for solving partial differential equations~\cite{FEniCS}.
A special feature of FEniCS is that it allows a straightforward definition of the variational formulation of differential equations in symbolic form using the Unified Form Language (UFL)~\cite{UFL}. Moreover, various types of finite elements can be used for spatial discretisation and numerous linear or nonlinear  solvers are available in FEniCS from third-party libraries, such as PETSc (Portable, Extensible Toolkit for Scientific Computing)~\cite{petsc-web-page, petsc-user-ref, petsc-efficient}.
Finally, FEniCS has been developed with high-performance computing in mind~\cite{FEniCS-book}. Hence, the FEDM code can be executed in parallel using MPI (message-passing interface), which can significantly speed up the solution of the problem on multicore architectures.

FEniCS has already been used successfully for the solution of plasma models~\cite{Hromadka-2016-ID5393}.
However, the FEDM code significantly extends the basic functionality of FEniCS so that especially large and complex plasma models can be implemented as easily as possible and solved efficiently in one (1D) or two (2D) spatial dimensions, noting that it is primarily designed for axisymmetric problems.
An automated definition of the balance equations on the basis of a prescribed species list and a reaction kinetics scheme was implemented to simplify the inclusion of extensive plasma chemistries involving a large number of reactions and species.
In order to cover a wide time range for simulations and overcome the equation stiffness problem that is common in plasma modelling, a user-defined time discretisation is realised by an algorithm based on a backward differentiation formula (BDF) with variable step size.
Finally, the code can solve equations using either a fully coupled or a segregated approach. 

The general features and use of the code are demonstrated by several examples, including
modelling of a time-of-flight experiment, a positive streamer in air at atmospheric pressure, and an abnormal glow discharge in argon at low pressure.
Using these examples, the verification of the code
is carried out by the method of exact solutions and benchmarking~\cite{Salari-2000-ID3784, Turner-2017-ID4135}. The mesh and time order-of-accuracy are determined as a rigorous verification of the code using the time-of-flight experiment. Streamer benchmark results from Ref.~\cite{Bagheri-2018-ID5240} are used as a consistency test for the implementation of the coupled approach in the present code. Furthermore, the streamer benchmark model is used to quantify the parallel performance of the FEDM code. The obtained speed-up is compared to the one obtained by the commercial software COMSOL Multiphysics\textsuperscript{\textregistered}~\cite{comsol} for the same test problem. Multi-node cluster calculations are additionally carried out as further performance tests. In order to illustrate the use of the code for further practical application cases, and to carry out further verification, results of the modelling of an abnormal glow discharge in argon at low pressure are presented and compared to results obtained using COMSOL Multiphysics\textsuperscript{\textregistered}.

The manuscript is organised as follows. In section 2, an overview of the governing model equations and corresponding boundary conditions is presented. The implementation of the model in FEniCS is described in the section 3. Section 4 illustrates the use of the code by performing three case studies, along with the verification of the code using the method of exact solutions and benchmarking. Section 5 examines the parallel performance of the code, and section 6 provides a brief summary.

\section{Governing equations and boundary conditions}
\label{model}
Regardless of the type of electric discharge being modelled, e.g.\ glow discharge at low pressure or streamer discharge at atmospheric pressure, the set of equations that needs to be solved in the framework of fluid-Poisson models is mostly the same. The set of equations implemented in the FEDM code includes balance equations for the particle number densities of the species considered in the model
\begin{equation}\label{eq:Continuity equation}
 \frac{\partial n_p}{\partial t} + \nabla \cdot \mathbf{\Gamma}_p=S_{p},
\end{equation}
where $n_p$ is the particle number density, $\mathbf{\Gamma}_p$ is the particle flux, $S_p$ is the source term describing the gain and loss of particles due to collision and radiation processes, and the index \textit{p} denotes the electrons, ions, and neutral species. The set of balance equations is coupled with Poisson's equation for the electric potential $\phi$
\begin{equation}\label{eq:Poisson's equation}
-\varepsilon_0 \varepsilon_\mathrm{r} \nabla^2 \phi = \rho\,,
\end{equation}
\noindent where $\rho=\sum_p q_{p} n_{p}$ is the space charge density, $q_p$ the charge of species \textit{p}, and $\varepsilon_\mathrm{r}$ and $\varepsilon_0$ denote the relative permittivity of the medium and the vacuum permittivity, respectively.
The particle fluxes in equation \eqref{eq:Continuity equation} are defined in a drift-diffusion approximation according to
\begin{equation}\label{eq:dda}
\mathbf{\Gamma}_\mathrm{e} =n_\mathrm{e} b_\mathrm{e} \nabla{\phi}  - \nabla(D_\mathrm{e} n_\mathrm{e})\,,
\end{equation}
\begin{equation}\label{eq:dda}
\mathbf{\Gamma}_{p} = -\textnormal{sgn}(q_p) n_p b_p \nabla{\phi}  - D_p \nabla n_p\,,
\end{equation}
\noindent where $b_p$ and $D_p$ are the mobility and the diffusion coefficient of the \mbox{\textit{p}-th} heavy particle species, respectively and $b_\mathrm{e}$ and $D_\mathrm{e}$ are the mobility and the diffusion coefficient of electrons. Note that placing the diffusion coefficient inside the gradient operator for electrons originates from the derivation of the drift-diffusion approximation (see \cite{Hagelaar-2005-ID2276, Grubert-2009-ID2551}) and is necessary to be consistent with the method of determination of the electron transport coefficients.
 The definition of the electron transport coefficients is crucial for the accuracy of a fluid model for non-thermal plasmas~\cite{Grubert-2009-ID2551,Becker-2017-ID4159,Baeva-2020-ID5434}.
To be able to apply these coefficients as function of the mean electron energy  $u_\mathrm{e}$ in accordance with the LMEA~\cite{Grubert-2009-ID2551}, the set of equations \eqref{eq:Continuity equation}--\eqref{eq:dda} is extended by the electron energy balance equation
\begin{equation}\label{eq:Energy balance}
 \frac{\partial w_\mathrm{e}}{\partial t} + \nabla \cdot \vb*{Q}_\mathrm{e} =-e_{0} \vb*{E} \cdot \mathbf{\Gamma}_\mathrm{e} + \widetilde{S}_\mathrm{e}\,.
\end{equation}
Here, $w_\mathrm{e}=n_\mathrm{e} u_\mathrm{e}$ is the electron energy density,
$\vb*{E}=-\nabla\phi$ is the electric field, $\widetilde{S}_\mathrm{e}$ is the energy source term describing the gain and loss of electron energy in collision processes, and $\vb*{Q}_\mathrm{e}$ denotes the electron energy flux given by
\begin{equation}\label{eq:Energy flux}
\vb*{Q}_\mathrm{e} = w_\mathrm{e} \tilde{b}_\mathrm{e} \nabla{\phi}  - \nabla(\tilde{D}_\mathrm{e} w_\mathrm{e})\,.
\end{equation}
The energy transport coefficients of electrons, $\tilde{b}_\mathrm{e}$ 
and $\tilde{D}_\mathrm{e}$, also depend on the mean electron energy. Note that solving the energy balance equation (\ref{eq:Energy balance}) is not required when using the LFA for determination of electron transport and rate coefficients. Both approaches (LMEA and LFA) are supported by the FEDM code.

In order to complete the model, an appropriate set of boundary conditions and initial conditions has to be specified. For Poisson's equation, Dirichlet boundary conditions specifying the applied voltage and the ground are generally applied at the electrodes, while homogeneous Neumann boundary conditions are applied at all other boundaries of the simulation domain. The following flux boundary conditions of the Robin type are applied for the heavy particle balance equations~\cite{Hagelaar-2000-ID1480, Becker-2013-ID3200}
\begin{equation}\label{eq:Particle flux for heavy particles}
\boldsymbol{\nu}  \cdot  \mathbf{\Gamma}_{p} = \frac{1-r_p}{1+r_p} \Big( |\textnormal{sgn}(q_p) b_p \boldsymbol{\nu} \cdot \vb*{E} n_p| + \frac{1}{2} v_{\mathrm{th}, p} n_p \Big)\,,
\end{equation}
where $\boldsymbol{\nu}$ is the outward normal on the boundary, $v_{\mathrm{th}, p}=\sqrt{\frac{8k_\mathrm{B} T_p}{\pi m_p}}$ denotes the thermal velocity of the species with mass $m_p$ and temperature $T_p$, $k_\mathrm{B}$ is the Boltzmann constant and $r_p$ denotes the reflection coefficient of the respective species. Similarly, the boundary conditions for the electrons read
\begin{eqnarray}
\label{eq:Particle flux for electrons bc}
\boldsymbol{\nu}  \cdot  \mathbf{\Gamma}_\mathrm{e} = \frac{1-r_\mathrm{e}}{1+r_\mathrm{e}} \Big( |b_\mathrm{e} \boldsymbol{\nu} \cdot \vb*{E} n_\mathrm{e}| + \frac{1}{2} v_\mathrm{th, e} n_\mathrm{e} \Big) \nonumber \\ - \frac{2}{1+r_\mathrm{e}} \gamma \sum_{i} \textnormal{max}(\boldsymbol{\nu} \cdot \mathbf{\Gamma}_i, 0)\,,\\
\label{eq:Energy flux for electrons bc}
\boldsymbol{\nu} \cdot \vb*{Q}_\mathrm{e} = \frac{1-r_\mathrm{e}}{1+r_\mathrm{e}} \Big( |\widetilde{b}_\mathrm{e} \boldsymbol{\nu} \cdot \vb*{E} w_\mathrm{e}| + \frac{2}{3} v_\mathrm{th, e} w_\mathrm{e} \Big) \nonumber \\ - \frac{2}{1+r_\mathrm{e}} u_\mathrm{e}^\gamma \gamma \sum_{i} \textnormal{max}(\boldsymbol{\nu} \cdot \mathbf{\Gamma}_i, 0)\,,
\end{eqnarray}
where
$v_\mathrm{th, e}=\sqrt{\frac{8k_\mathrm{B}T_\mathrm{e}}{\pi m_\mathrm{e}}}$ and $k_\mathrm{B} T_\mathrm{e}= 2u_\mathrm{e}/3$.
The second term on the right hand side of equations~\eqref{eq:Particle flux for electrons bc} and \eqref{eq:Energy flux for electrons bc} describes the secondary emission of electrons from the boundaries due to particle bombardment with secondary electron emission coefficient $\gamma$ and the mean energy of emitted electrons $u_\mathrm{e}^\gamma$.
These boundary conditions may be changed to account for further physical effects, such as the accumulation of surface charges on dielectrics and photoemission.

Finally, appropriate initial conditions need to be set before solving the problem. Usually, quasi-neutral conditions with a uniform density for all species are assumed.

\section{Code implementation}

The main aim of the 
FEDM code is to 
simplify the implementation of the governing equations, 
i.e.\ equations~\eqref{eq:Continuity equation}--\eqref{eq:Energy flux} with boundary conditions~\eqref{eq:Particle flux for heavy particles}--\eqref{eq:Energy flux for electrons bc}, in FEniCS.
For simpler problems, this can be done manually, but the implementation of more complex plasma models can be considerably supported using automation techniques~\cite{Jovanovic-2021-ID5864}.
Hence, it is explained first how the equations can be implemented  natively in FEniCS and then the automation procedure introduced by the FEDM code is described.

\subsection{Variational problem definition in FEniCS}
In order to solve the equations in FEniCS, the governing equations need to be defined in variational (weak) form.
This form
is obtained by multiplying the respective general (strong) form
with proper test function $v\in V$, integration over the given solution domain $\mathit{\Omega}$, and integrating the flux term by parts~\cite{Zienkiewicz,FEniCS-book}. Here, $V$ denotes a suitable function space~\cite{Zienkiewicz}, which might be different for the individual equations. With this, the system of Poisson's equation, the balance equations for the particle number densities and the electron energy balance equation reads~\cite{Becker-2009-ID2678}

\begin{align}
&\int_{\mathit{\Omega}}^{} \left(\varepsilon_{0} \varepsilon_\mathrm{r} \nabla \phi \cdot \nabla v
- \rho  v \right) \mathrm{d} \mathit{\Omega} \nonumber \\
&\quad- \int_{\partial \mathit{\Omega}}^{} \varepsilon_{0} \varepsilon_\mathrm{r} \boldsymbol{\nu}\cdot \nabla \phi\, v \mathrm{d}s = 0,
\quad\forall v\in V_\phi\,,\label{eq:poisson-weak}\\
&\int_{\mathit{\Omega}}^{} \left( \frac{\partial n_p}{\partial t} v
- \mathbf{\Gamma}_p \cdot \nabla v
- S_{p} v \right)\mathrm{d} \mathit{\Omega} \nonumber\\
&\quad+ \int_{\partial \mathit{\Omega}}^{} \boldsymbol{\nu}\cdot \mathbf{\Gamma}_p v\, \mathrm{d}s = 0,
\quad\forall v\in V_p\,,\label{eq:ni-weak}\\
&\int_{\mathit{\Omega}}^{} \left( \frac{\partial w_\mathrm{e}}{\partial t} v
- \vb*{Q}_\mathrm{e} \cdot \nabla v
+ e_{0} \vb*{E} \cdot \mathbf{\Gamma}_\mathrm{e} v
- \tilde{S}_\mathrm{e} v \right)\mathrm{d} \mathit{\Omega} \nonumber\\
&\quad+ \int_{\partial \mathit{\Omega}}^{} \boldsymbol{\nu}\cdot \vb*{Q}_\mathrm{e} v\, \mathrm{d}s = 0,
\quad\forall v\in \widetilde{V}_\mathrm{e}\,,\label{eq:we-weak}
\end{align}
where $\boldsymbol{\nu}$ is the outward normal to the boundary $\partial \mathit{\Omega}$ and $\mathrm{d}s$ is the surface area element.
Flux boundary conditions are introduced by replacing the corresponding flux terms in the boundary integrals. The Dirichlet boundary conditions for the electric potential are taken into account via proper definition of the function space $V_\phi$~\cite{Becker-2009-ID2678}.
It should be noted that the equations must be discretised in time by applying a time discretisation method, such as the backward differentiation formula, which is not natively supported in FEniCS.
The variational form is then symbolically defined using UFL, and automatically discretised by FEniCS. For this, proper discrete function spaces must be chosen, which is done by the choice of finite elements~\cite{FEniCS-book} and the used mesh. Note that FEniCS supports the use of various types of elements, such as Lagrange, Brezzi-Douglas-Marini, Raviart-Thomas and others~\cite{FEMtable}, and provides the possibility to use the discontinuous Galerkin method for discretisation of differential equations.
Furthermore, the mesh can be generated either by using a built-in FEniCS function (which is limited to structured triangular meshes) or by importing an externally generated mesh from \texttt{xml} or \texttt{xdmf} files.
By discretising the problem, a system of nonlinear equations is obtained. This system of equations can be solved using external libraries, such as PETSc.

\begin{figure*}[!h]
\centering
\includegraphics[scale = 0.8]{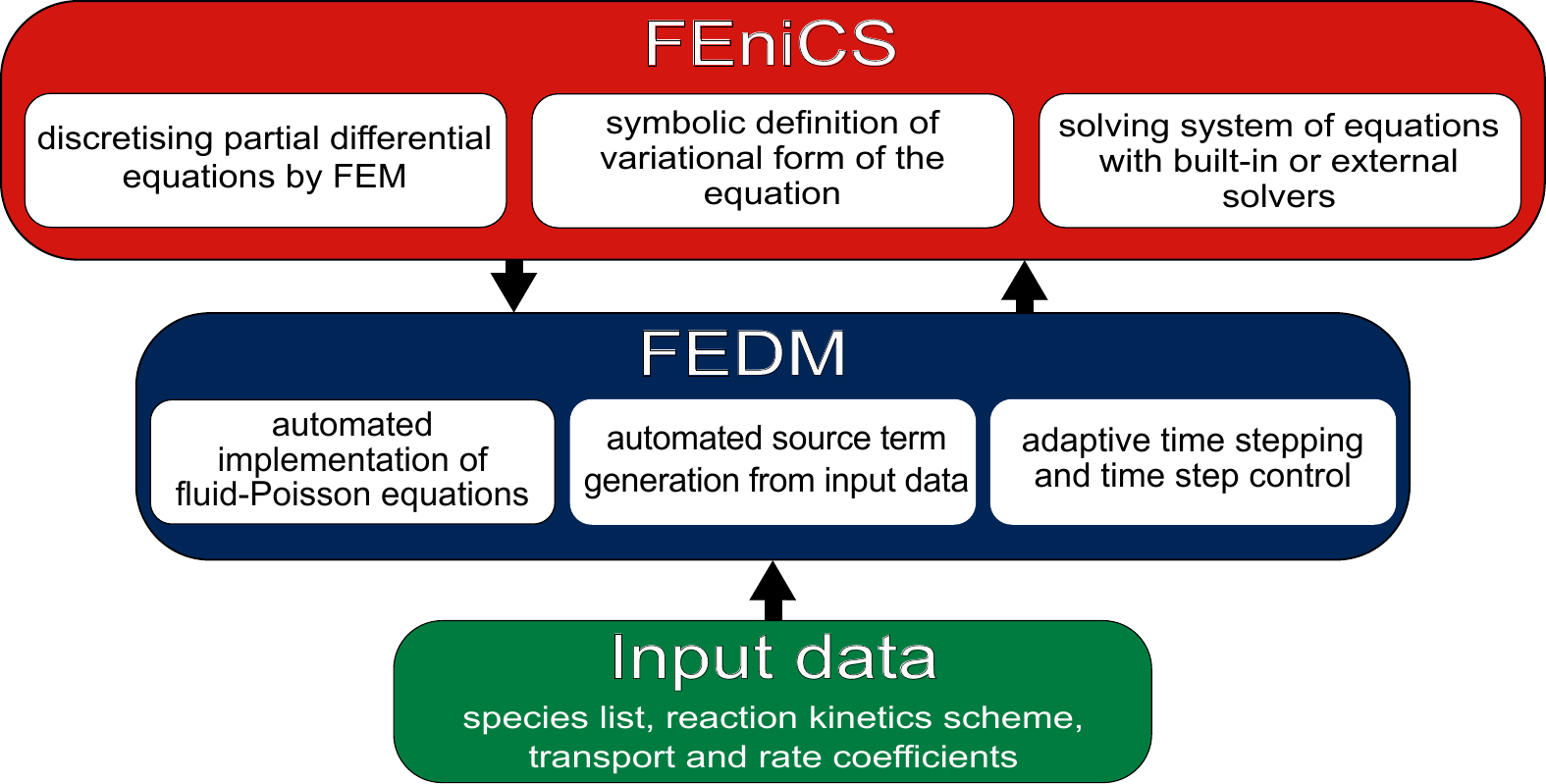}
\caption{Diagram of the features of the FEDM code and its interconnection with the input data and FEniCS.}
\label{fig:diagram_FEDM}
\end{figure*}

\subsection{FEDM code}

For problems involving a few species only, the described procedure can be easily performed manually.
Challenges arise when a large number of species needs to be taken into account, i.e.\ when many equations need to be solved simultaneously.
In that case, the manual definition of test and trial functions, variational forms of particle balance equations, transport coefficients and source terms (which can contain numerous chemical reactions, in some cases hundreds or more) becomes a time-consuming, tedious and error-prone process.
Moreover, the lack of support for time discretisation and adaptive time stepping by FEniCS leads to the necessity to implement them manually for all time-dependent equations.

In order to overcome these challenges, the FEDM code introduces functions for an automated definition of the variational form of the balance equations for an arbitrary number of species prescribed in a species list (cf.~Figure \ref{fig:diagram_FEDM}).
In addition, the code provides functions for reading in a reaction kinetic scheme and automated definition of the source terms based on the given scheme.
This can drastically reduce the implementation time of a new model and also limit the possibility of errors, which can occur when doing this process manually for each of the equations.
The required input data, such as the number of species and their respective properties (mass and charge), is stored in the configuration files.
Furthermore, the respective transport and rate coefficients can be stored as constant values, in a form of functions (written as Python code) or as tabulated data in separate input files.
In the latter case, the data is imported into the code as look-up tables.
The reaction kinetic scheme determining the source terms of the balance equations is
used to automatise the definition of the rates and source terms in the following way.
First, the rates are calculated as
\begin{equation}\label{eq:rate}
R_j = k_j \prod_{p=1}^{\mathrm{N}_\mathrm{s}} n_p^{\beta_{pj}}\,,
\end{equation}
where $\mathrm{N}_\mathrm{s}$ is the number of species, $n_p$ is the number density of the \mbox{$p$-th} species, and $\beta_{pj}$ is the partial reaction order of species $p$ in reaction $j$. The source terms are then defined as
\begin{equation}\label{eq:source_term}
S_p = \sum_{j=1}^{\mathrm{N}_\mathrm{r}} (G_{pj} - L_{pj}) R_j\,,
\end{equation}
where $\mathrm{N}_\mathrm{r}$ is the number of reactions and the respective matrix elements $G_{pj}$ and $L_{pj}$ contain the stoichiometric coefficients for given species $p$ in reaction $j$. Similarly, when LMEA is used, the energy source term is defined as
\begin{equation}\label{eq:energy_density_source_term}
\widetilde{S}_\mathrm{e} = \sum_{j=1}^{\mathrm{N}_\mathrm{r}} \Delta \varepsilon_j R_j + \widetilde{S}_\mathrm{el} + \widetilde{S}_\mathrm{att} + \widetilde{S}_\mathrm{rec}\,,
\end{equation}
where $\Delta \varepsilon_j$ is the energy loss (or gain) for the \mbox{$j$-th} reaction and the terms $\widetilde{S}_\mathrm{el}$, $\widetilde{S}_\mathrm{att}$ and $\widetilde{S}_\mathrm{rec}$
describe the energy change in elastic collisions, electron attachment and electron-ion-recombination processes, respectively~\cite{Gnybida-2009-ID2570, Ponduri-2016-ID3865}.

\begin{figure*}[!htbp]
\centering
\includegraphics[scale = 0.8]{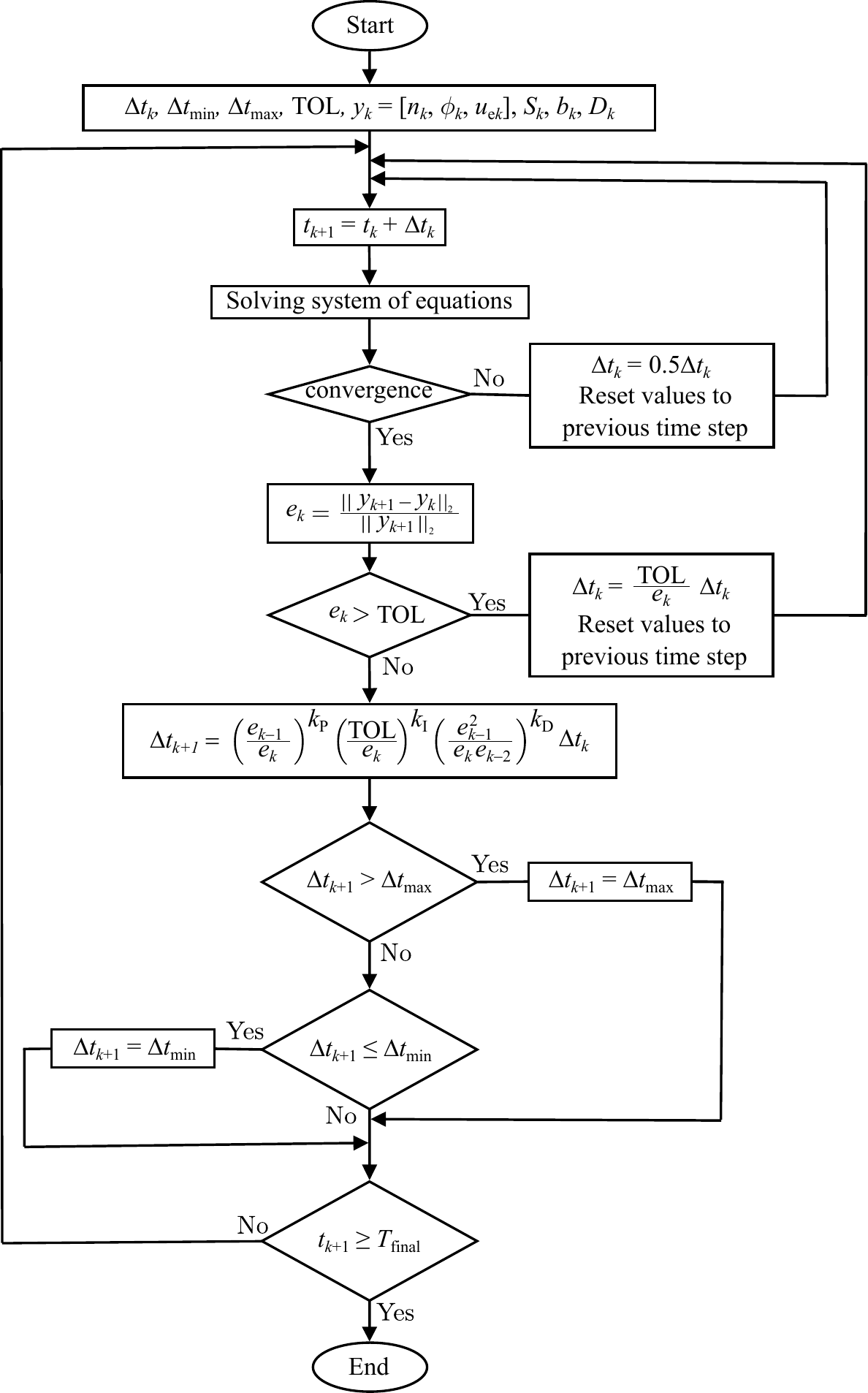}
\caption{Flow chart describing the adaptive time-stepping procedure. The problem is solved with initial time step size $\Delta t_k$ and if the convergence is reached and the prescribed tolerance is satisfied, the new time step $\Delta t_{k+1}$ is calculated using the PID controller; otherwise, the time step is halved and the calculations are repeated. If the new time step is greater than a prescribed time step size $\Delta t_\mathrm{max}$, or lower than a prescribed time step size  $\Delta t_\mathrm{min}$, it takes the respective minimum or maximum value and the calculations continue.
The calculation ends when the end time $T_\mathrm{final}$ is reached.}
\label{fig:flow_chart}
\end{figure*}

The very different time scales of the various reaction processes included in the model can lead to the occurrence of a system of stiff differential equations, which constitutes another challenge to overcome.
This is tackled by implementation of an implicit time discretisation method suitable for stiff equations. Moreover, the time range that the model needs to cover sometimes spans from picoseconds to tens of microseconds or even seconds, i.e.\ the use of an adaptive time-stepping method is required.
The FEDM implements a variable step-size BDF method of second order~\cite{HairrerNonStiff, Celaya-2014-ID5395} for time discretisation of the equations, which reads
\begin{equation}
y_{k+1} - \frac{(1+\omega_{k})^2}{1+2\omega_{k}} y_{k} + \frac{\omega_{k}^2}{1+2\omega_{k}} y_{k-1}  = \Delta t_{k} \frac{1+\omega_{k}}{1+2\omega_{k}} f_{k+1}\,.
\end{equation}
Here, $\Delta t_{k}$ and $y_{k}$ are the time step size and the value of the unknown in time step $k$, respectively, and $\omega_{k} = \Delta t_{k}/ \Delta t_{k-1} $. 
This time discretisation method and its specific order were chosen for efficiency reasons since they provide good accuracy and stability with a small memory footprint. Higher-order BDF schemes could also be implemented, with the requirement of introducing order control.

The adaptive time step control is realised by means of an evolutionary proportional-integral-derivative (PID)
controller~\cite{Zhuang-1994-ID5401, tsm_course}.
The implemented method is based on a local truncation error analysis and consists of calculating the relative change of the unknown variable as a criterion for the time step refinement. The procedure consists of the following steps. First, the value of the unknown function $y_{k+1}$ is calculated at the time $t_{k+1}$. Then, the relative change of the unknown variable $e_k$ between two time steps  is calculated as
\begin{equation}\label{eq:error}
e_{k} = \frac{\norm{y_{k+1}-y_{k}}_2}{\norm{y_{k+1}}_2 }\,,
\end{equation}
where $y_{k+1}$ and $y_{k}$ are values of the unknown in time steps $k+1$ and $k$, respectively. 
If the relative change is larger than a tolerance or if the nonlinear solver fails, calculations are repeated with a smaller time step size; otherwise, the new time step size is determined using the formula~\cite{tsm_course}
\begin{equation}\label{eq:ATS}
\Delta t_{k+1} = \bigg(\frac{e_{k-1}}{e_k} \bigg)^{k_\mathrm{P}}  \bigg(\frac{\mathrm{TOL}}{e_k}\bigg)^{k_\mathrm{I}}  \bigg(\frac{e_{k-1}^2}{e_k e_{k-2}}\bigg)^{k_\mathrm{D}} \Delta t_{k},
\end{equation}
where $\mathrm{TOL}$ is the prescribed tolerance, and $k_\mathrm{P}$, $k_\mathrm{I}$ and $k_\mathrm{D}$ are empirically determined constants. In order to limit an uncontrolled increase of the time step size leading to the failure of the nonlinear solver, minimum and maximum values for the time step size can be prescribed. It should be pointed out that the time stepping is significantly affected by the convergence of the solution as it depends directly on the local truncation error. Using a finer mesh to resolve steep gradients can help to achieve better convergence and smaller errors, allowing larger time steps. The flow chart of implementation of the evolutionary PID controller for adaptive time stepping in the FEDM code is presented in Figure~\ref{fig:flow_chart}.

After setting up the variables and the source terms, the system of variational equations~\eqref{eq:poisson-weak}--\eqref{eq:we-weak} is automatically set up by the FEDM code, where the number of balance equations~\eqref{eq:ni-weak} equals the number of species considered in the model.

Further features of the FEDM code ensure the robustness of the solution procedure.
It provides the option to solve the balance equations~\eqref{eq:ni-weak} and~\eqref{eq:we-weak} in logarithmic form by replacing the solution variable with its natural logarithm, e.g.\ $n^*_p = \ln(n_p)$. One benefit of this approach is that large-scale variations are avoided, as are negative values in the solution~\cite{Welland2014}.
Moreover, it is well-known that the way of treating the source terms in Poisson's equation and the electron energy balance equation 
strongly affects the maximum time step size for which a stable solution can be obtained~\cite{LIN20121225, Ventzek-1994-ID5407, HAGELAAR20001}.
Namely, if the source term in Poisson's equation is treated explicitly, the size of the time step is limited by the dielectric relaxation time~\cite{LIN20121225, Ventzek-1994-ID5407}.
The same holds for the source term in the electron energy balance equation when the transport and rate coefficients are calculated based on the mean electron energy or electric field values of the previous time step~\cite{HAGELAAR20001}.
To overcome the problem of small time step sizes, a semi-implicit treatment of these source terms has been suggested~\cite{Ventzek-1994-ID5407, HAGELAAR20001}. However, a fully coupled solution of the whole system of partial differential equations, where the variational forms of the equations are summed up and all equations are solved simultaneously, allows larger time step sizes and provides higher robustness in comparison with a segregated solution of the equations, where each equation is solved separately.
FEniCS supports both the segregated and fully coupled approach. In the FEDM code, all equations are solved in the fully coupled manner by default (although in some cases transport and rate coefficients are calculated based on the values from the previous time step using the aforementioned semi-implicit approaches). The fully coupled approach requires solving a nonlinear system of equations at each time step. For solution of this nonlinear system the FEDM code provides access to the SNES (Scalable Nonlinear Equations Solvers) component of the PETSc library, which is part of FEniCS. By default, FEDM uses the Newton-based solver provided by SNES  and the direct MUltifrontal Massively Parallel sparse direct Solver (MUMPS)~\cite{mumps-web-page} to solve the linear equation system during each Newton iteration. Note that PETSc also provides other nonlinear and linear solvers that can be used by the FEDM code. 

The FEDM code was designed for the analysis of non-thermal discharges. The focus on simple case studies presented in the following section is mostly for verification purposes. The FEDM code can be easily adapted to model different types of discharges at low and atmospheric pressure in plane-parallel configuration or geometries containing curved boundaries (see, for example~\cite{Jovanovic-FEniCS2021}). The equations can be solved in spatially 1D (Cartesian and polar coordinates) and 2D (Cartesian and cylindrical coordinates) domains\footnote{FEniCS generally solves the equations in 3D Cartesian coordinates noting that the implementation and solution of a 3D plasma model remains a larger challenge.}. In addition to basic FEniCS functionality that uses the mesh function to read the physical tag of the boundaries, the FEDM code introduces custom marking of arbitrary boundaries, which is important for properly imposing Neumann or Robin boundary conditions, e.g.\ on internal interfaces. For discharges with multiple domains (e.g.\ dielectric barrier discharges with plasma, dielectric and the interface between them), the mixed-dimensional approach is available via FEniCS~\cite{Daversin-Catty-2021-ID6140}. Additional functionalities can be added by FEniCS or external libraries. For instance, the present code version does not consider neutral gas flow, gas heating or photoionisation. The additional equations (i.e.\ Navier-Stokes equations or Helmholtz equation for photoionisation~\cite{Bourdon-2007-ID3277}) can be manually implemented and solved using FEniCS. Although the FEDM code does not account for the gas flow, other open-source codes designed for solving the Navier-Stokes equations are available in FEniCS (see, for instance,~\cite{Mortensen-2015-ID6141}), which could be coupled with the FEDM code. Finally, it should be highlighted that besides simplifying the modelling procedure, the Python interface used by the FEDM code provides direct access to widely available data science tools, making post-processing of the results convenient.

\section{Code verification and application examples}
This section presents three case studies to verify the FEDM code and to illustrate its practical application.
First, a time-of-flight experiment is modelled, where the evolution of the electron number density is calculated and compared to the analytically derived exact solution. In addition to this consistency test, a rigorous verification of the code is carried out by performing order-of-accuracy studies~\cite{Salari-2000-ID3784} for the space and time discretisation, on the basis of which the order of accuracy is determined.
In the second case study, the positive streamer benchmark in air at atmospheric pressure proposed by Bagheri~\textit{et al.}~\cite{Bagheri-2018-ID5240} is used for a further consistency check of the FEDM code. Although less rigorous than the method of exact solutions, benchmarking is useful for verifying more complicated codes, which are employed to simulate models without a known analytical solution. The third case study presents all features provided by the FEDM code. Here, an abnormal glow discharge in argon at low pressure is modelled, and FEDM results are compared to results obtained by the commercial software package COMSOL Multiphysics\textsuperscript{\textregistered} for further benchmarking.

\subsection{Time-of-flight experiment}
\label{sec:tof-experiment}
\begin{figure}[!t]
	\centering
	\includegraphics[width = 8cm]{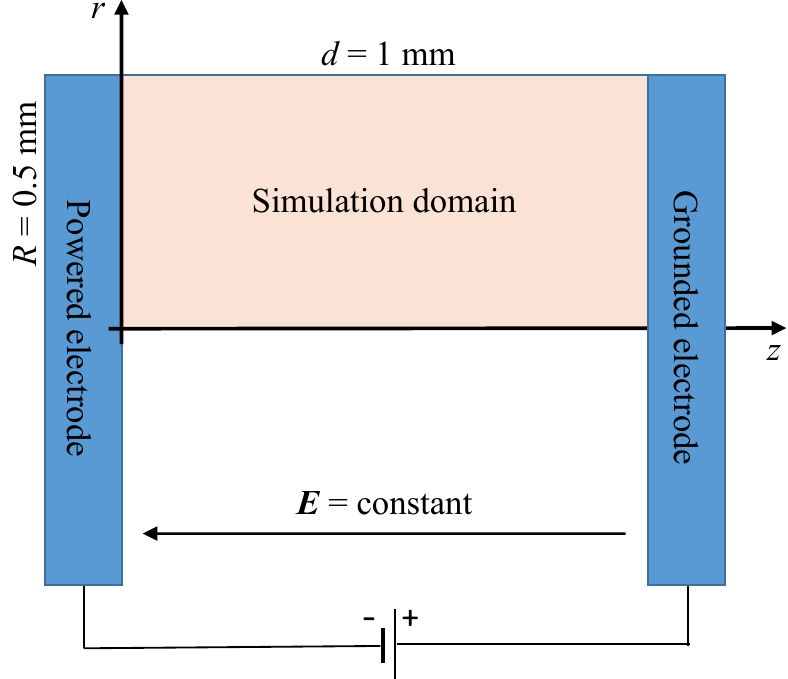}
	\caption{Simulation domain for the time-of-flight experiment. Due to axial symmetry, the problem was solved in cylindrical coordinates. The domain was discretised using a structured triangular mesh with the number of elements ranging from $10^4$ to $4\times10^6$.}
	\label{fig:geometry}
\end{figure}

In the first case study, the FEDM code was used to model the spatiotemporal evolution of an electron cloud in a time-of-flight experiment using the plane-parallel axisymmetric geometry shown in Figure~\ref{fig:geometry}.
A constant axial electric field was applied between the powered and the grounded electrode separated by a gap of $d=1$\,mm and the radial component of the electric field was zero. An electrode radius of $R=0.5$\,mm was assumed.
Starting from initial conditions for $t=t_0>0$ according to
\begin{equation}\label{eq:tof-sol}
n_\mathrm{e}(r, z, t) = (4 \pi D_\mathrm{e} t)^{-3/2}\, \mathrm{e}^{-\frac{(z - v_\mathrm{e} t)^2+r^2}{4 D_\mathrm{e} t}+(\alpha - \eta)v_\mathrm{e}t}\,,
\end{equation}
the spatiotemporal evolution of the electron number density $n_\mathrm{e}$ can be simulated by solving the continuity equation~\eqref{eq:Continuity equation} for electrons with the flux
\begin{equation}\label{eq:tofflux}
\vb*{\Gamma}_\mathrm{e} = - \nabla (D_\mathrm{e} n_\mathrm{e}) + n_\mathrm{e}\vb*{v}_\mathrm{d}
\end{equation}
and the source term
\begin{equation}\label{eq:tofsource}
S_\mathrm{e} = (\alpha - \eta) n_\mathrm{e} v_\mathrm{d}\,.
\end{equation}
Here, $\alpha$ and $\eta$ are Townsend's ionisation and attachment coefficients, respectively, and the electron drift velocity $\vb*{v}_\mathrm{d}$ has the axial component ${v_\mathrm{d}}_z = v_\mathrm{d}$, while its radial component is zero. Constant coefficients calculated for the given constant electric field were used in the example to be able to compare the modelling result with the available analytical solution.
For times $t\geq t_0$ the exact solution of the continuity equation of electrons is given by equation~\eqref{eq:tof-sol}~\cite{RaizerGDP, Blevin-1984-ID2939} and can be used to verify the FEDM code by means of the method of exact solutions.

For the numerical solution of the given test problem using the FEDM code, equation (\ref{eq:Continuity equation}) with flux~\eqref{eq:tofflux} and source term~\eqref{eq:tofsource} was discretised in space using linear Lagrange elements and the time discretisation was performed by the second-order BDF method using a constant time step size.
The electric field strength was set to $|\vb*{E}| = 3.55$\,MV/m. 
Assuming synthetic air at atmospheric pressure and a gas temperature of 300 K as background gas, the values of the drift velocity, diffusion coefficient and effective ionisation coefficient at these conditions are $v_\mathrm{d} = 1.7 \times 10^5$\,m/s, $D_\mathrm{e} = 0.12$\,m$^2$/s and $\bar{\alpha} = \alpha - \eta = 5009.5$\,m$^{-1}$, respectively~\cite{Bagheri-2018-ID5240}.
To avoid the distortion of the external electric field and 
influence by the boundaries,
the simulations were carried out in the time interval from $t_0 = 2$\,ns to $t=4$\,ns, during which the number density remains low enough and the spatial profile of the electron cloud stays far away from the boundaries.
Note that homogeneous Neumann boundary conditions were applied at both sides. The obtained system of nonlinear equations was solved using the Newton method provided by PETSc SNES. The number of nonlinear iterations was limited to 50 and the relative tolerance was fixed to $10^{-10}$ for the nonlinear solver. MUMPS was used for solving the linear system of equations in each iteration of the nonlinear solver.

\begin{figure}[!htbp]
\centering
\includegraphics[scale=1.0]{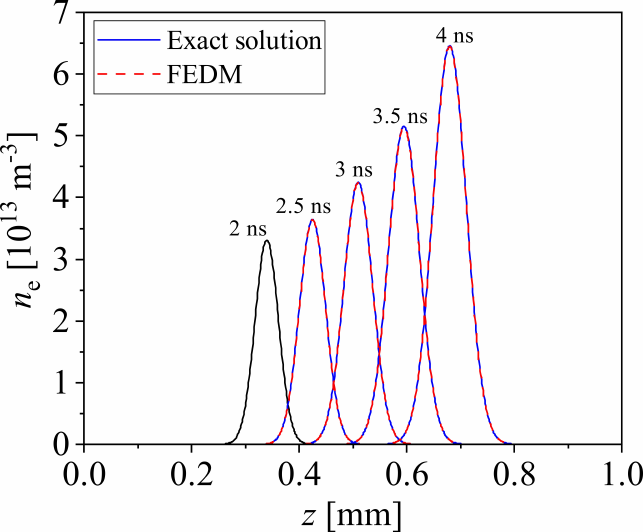}
\caption{Temporal evolution of the electron number density profile at the symmetry axis. The simulations started at $t_0 = 2$\,ns and were carried out with the constant time step size $\Delta t = 1 \, \mathrm{ps}$ using a mesh consisting of $100000$ elements.}
\label{fig:MESprofiles}
\end{figure}

First, a consistency test was performed by comparing the numerical solution to the exact one. The constant time step size $\Delta t = 1 \, \mathrm{ps}$ and a mesh with approx.\ 100000 elements was used for this calculation.
The resulting evolution of the electron number density along the symmetry axis is presented in Figure~\ref{fig:MESprofiles}, showing excellent agreement between the numerical and the exact solution at all times.

\begin{figure}[!htbp]
\centering
\includegraphics[scale=1.0]{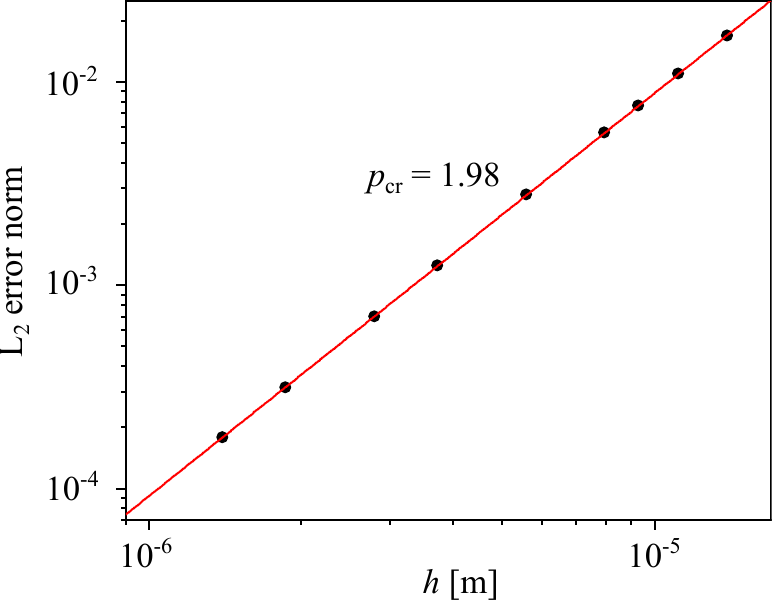}
\caption{Error of the numerical solution in $L_2$ norm in dependence on the element size $h$ for the time $t = 3$\,ns and time step size $\Delta t = 0.02 \, \mathrm{ps}$.}
\label{fig:meshOA}
\end{figure}

For proper code verification, more rigorous studies such as space and time order-of-accuracy tests are needed.
With this, it can be demonstrated that the equations are solved to the theoretical order of accuracy of the respective discretisation method~\cite{Salari-2000-ID3784}.
For this purpose, the $L_2$ error norm representing the distance between the exact and numerical solution, i.e.\ $\norm{y_\mathrm{exact}-y_\mathrm{numerical}}_2$ was calculated for different levels of refinements for the spatial mesh and time step size.
The discretisation error obtained from the spatial order-of-accuracy test is given as a function of the mesh element size $h$, which is equal to the cell diameter (circumradius of the triangle).
The spatial discretisation error was assumed to be proportional to the mesh size as $E_i = Ch_i^{p_\mathrm{cr}}$, where $C$ is a constant, $h_i$ is the mesh size for the 
\mbox{$i$-th} level of refinement, and $p_\mathrm{cr}$ is the convergence rate~\cite{Salari-2000-ID3784}.
Structured meshes with different degree of refinement with the number of elements ranging between 10000 and approx.\ 4 million were used to determine the spatial order of accuracy.
The calculations were carried out up to \mbox{$t = 3 \, \mathrm{ns}$}, where the electron density profile is far from both boundaries (cf. Figure~\ref{fig:MESprofiles}), using a constant time step size of \mbox{$\Delta t =  0.02 \, \mathrm{ps}$}.
The obtained results for the $L_2$ error norm for different levels of mesh refinement are presented in Figure~\ref{fig:meshOA}.
The convergence rate $p_\mathrm{cr} = 1.98$ was determined from the slope of a linear fit of the calculated data points.
This value agrees very well with the theoretical second order accuracy and verifies the implementation of the spatial discretisation method. 

The same order-of-accuracy test was performed for the time discretisation. To reduce the calculation time, the convergence rate was determined from the discretisation errors calculated for two subsequent refinement levels for the time step size according to
\begin{equation}
p_\mathrm{cr} = \frac{\textrm{log}_{10}(E_1/E_2)}{\textrm{log}_{10}(\Delta t_1/\Delta t_2)}\,.
\end{equation}
Here, the discretisation errors $E_1$ and $E_2$ represent the $L_2$ error norms for the numerical solutions obtained  for two different time step sizes $\Delta t_1$ and $\Delta t_2$.
A fine mesh with 500000 Lagrange elements of third order was used to determine the order of accuracy regarding the time discretisation.
The time step sizes $\Delta t_1 =  5 \, \mathrm{ps}$ and $\Delta t_2 =  4 \, \mathrm{ps}$ were employed to obtain the results in a reasonable time. The time convergence rate obtained in this way was equal to $p_\mathrm{cr}=1.99$, tending to the theoretical second order accuracy.

In summary, the results obtained for the spatial and temporal order-of-accuracy tests convincingly verify the FEDM code regarding the solution of the continuity equation of electrons at conditions of a time-of-flight experiment in air.

\begin{figure*}[!h]
	\centering
	\includegraphics[width=.95\linewidth]{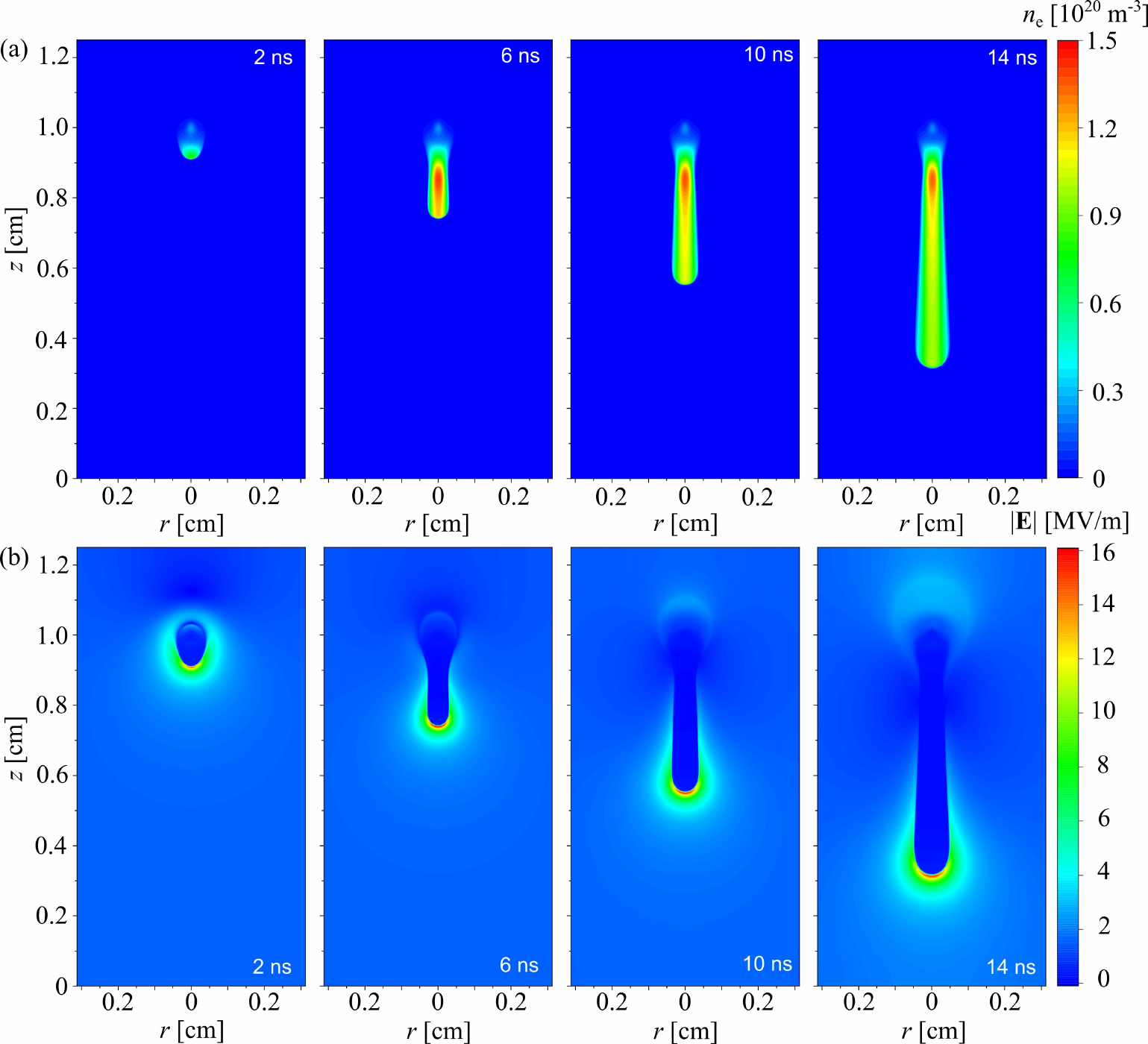}
	\caption{Temporal evolution of the spatial profiles of (a) the number density of electrons and (b) the electric field simulated up to 14\,ns applying the FEDM code for the streamer benchmark case.}
	\label{fig:profiles2Dt}
\end{figure*}

\subsection{Streamer benchmark case}
\label{sec:streamer-benchmark}
The determination of the analytical solution for the full set of nonlinearly coupled partial differential equations is hardly possible and achievable only for simplified cases.
Therefore, an alternative method has to be applied in order to test the implementation of the model and the accuracy of the code for real conditions.
In that case, a comparison with benchmark data being calculated by already established modelling codes represents a useful means for code verification.
The comparison of six streamer simulations codes by Bagheri~\textit{et al.}~\cite{Bagheri-2018-ID5240} was used to benchmark the FEDM code.
The considered streamer benchmark case describes the propagation of a positive streamer in synthetic air at ambient pressure and a gas temperature of 300\,K. It should be noted that test case~1 of Ref.~\cite{Bagheri-2018-ID5240} was selected for the comparison, where a relatively high level of background ionisation is considered and photoionisation is neglected.
Such conditions are typical of streamer breakdown in dielectric barrier discharges~\cite{Hoft-2016-ID5928}, which represents a reasonable application scenario for the FEDM code.

In accordance with the given benchmark model, the continuity equations~\eqref{eq:Continuity equation} were solved for electrons and positive ions, where the diffusion coefficient of electrons was placed outside the gradient operator and ions are considered to be immobile, i.e.\ the ion flux was set to zero.
The same mobility, diffusion coefficient and effective ionisation coefficient as in Ref.~\cite{Bagheri-2018-ID5240} were used for electrons (i.e.\ LFA was used), and the ionisation source term~\eqref{eq:source_term} for electrons and ions was defined as
\begin{equation}
	S_\mathrm{e} = S_\mathrm{i} = \bar{\alpha} b_\mathrm{e} |\vb*{E}| n_\mathrm{e}\,.
\end{equation}
The continuity equations were solved together with Poisson's equation~\eqref{eq:Poisson's equation} in a self-consistent manner. 
Using the same geometry as in figure 3 and applying the discharge parameters given in Ref.~\cite{Bagheri-2018-ID5240}, the plan-parallel electrodes were set to be $d = 1.25\,\mathrm{cm}$ apart, assuming an electrode radius of $R=1.25\,\mathrm{cm}$.
The constant applied voltage $U=-18.75$\,kV was applied to the powered electrode. This corresponds to a background electric field in $z$ direction of $E_0 = 15\,\mathrm{kV/cm}$. This electric field is being far below the breakdown field.
In order to locally enhance the electric field to the values above the breakdown threshold, a Gaussian seed of positive ions was introduced at the symmetry axis near the anode (starting point of the positive streamer).
Further details of the setup are described in Ref.~\cite{Bagheri-2018-ID5240}.

Benchmark calculations using the FEDM code were carried out using a triangular mesh with approx.\ 800000 elements, which was generated by means of gmsh~\cite{Geuzaine-2009-ID5399}.
The mesh was refined in the narrow streamer propagation region along the symmetry axis to fully resolve the steep gradients in front of the streamer head.
The minimum element size (length of the triangles) was set to the order of one micrometer and linear Lagrange elements were used for the spatial discretisation.
Furthermore, a variable step-size BDF method of second order with a maximum time step size of $\Delta t_\mathrm{max} = 5 \, \mathrm{ps}$ was used for the time discretisation. The electron number density was used for error control of the time-stepping procedure in this case study.
Note that the continuity equations were implemented in logarithmic form in order to improve the robustness of the calculations and to reduce the number of elements required for the numerical solution.

The system of nonlinear equations was solved using the nonlinear solver from the PETSc SNES library, where the direct solver MUMPS was used to solve the linear equation system during each iteration of the nonlinear solver.
The relative tolerance $\mathrm{TOL} = 10^{-3}$ was employed as a termination criterion for the iterations during each time step. The relative tolerance of the nonlinear solver was set to  $10^{-4}$, allowing high accuracy with a low number of nonlinear iteration steps.  The calculations were performed on a compute node with 16 physical cores and took about 5 hours.

Figure \ref{fig:profiles2Dt} shows the modelling results of the electron density (top) and the electric field strength (bottom).
It can be seen that a streamer starts to propagate from the seed point near the anode (top electrode) towards the cathode (bottom electrode) in accordance with the results presented in Ref.~\cite{Bagheri-2018-ID5240}.

\begin{figure*}[!t]
	\centering
	\includegraphics[height = 6cm]{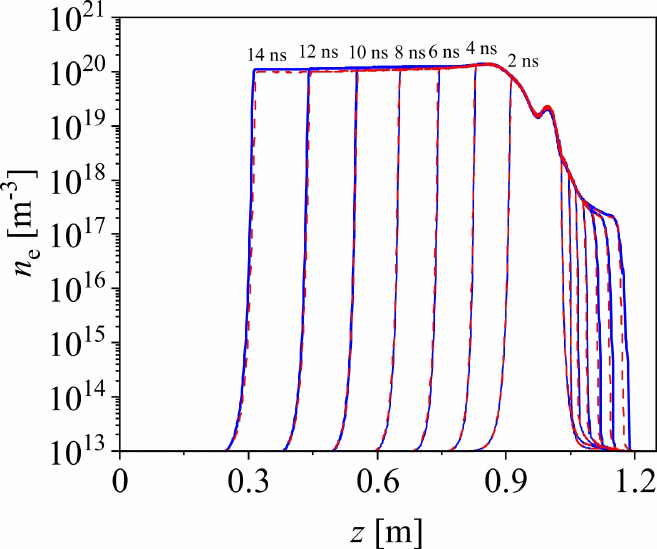}
	\includegraphics[height = 6cm]{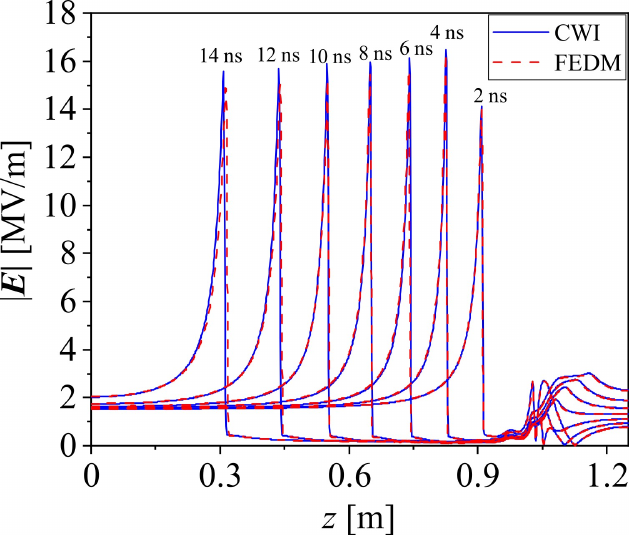}
	\caption{Comparison of (a) the electron number density and (b) the electric field profiles along the symmetry axis in the time range between 2 and 14\,ns calculated using the FEDM code (dashed red line) and the CWI benchmark data (solid blue line).
	}
	\label{fig:axial_profiles}
\end{figure*}

\begin{figure*}[!t]
	\centering
	\includegraphics[scale=0.9]{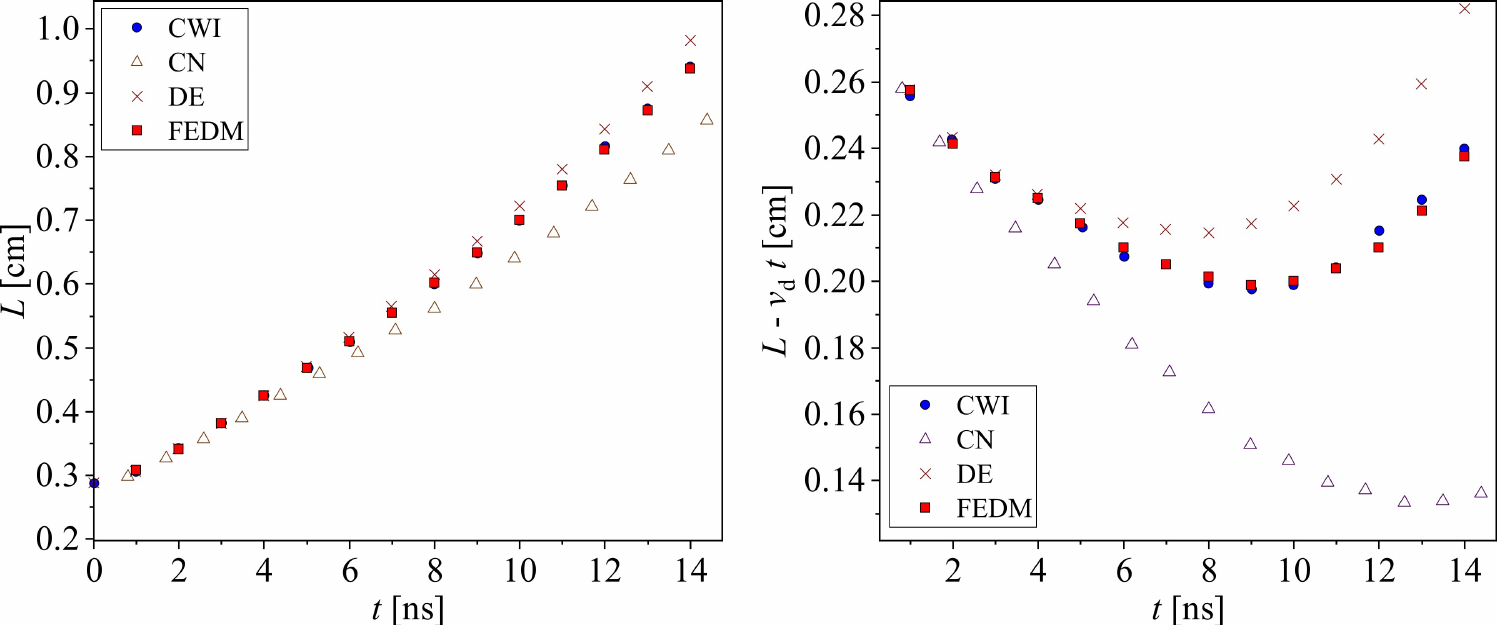}
	\caption{Comparison of (a) streamer length and (b) streamer length with subtracted $v_\mathrm{d} t$  as a function of time calculated using the FEDM  code with the results of several groups reported in~\cite{Bagheri-2018-ID5240}.}
	\label{fig:streamer length}
\end{figure*}

For a more rigorous comparison and benchmarking of the FEDM code, Figure~\ref{fig:axial_profiles} directly compares the results obtained by the FEDM code with the published data of the CWI group.
Figure~\ref{fig:axial_profiles}\,(a) shows the electron density along the symmetry axis for different times during the streamer propagation and Figure~\ref{fig:axial_profiles}\,(b) represents the axial electric field strength for the same instants.
The obtained agreement between the present results and those obtained by the CWI group is very good. This further confirms that the FEDM code provides reliable results, especially when considering the wide range of simulation results of the codes involved in the benchmark study of Bagheri~\textit{et al.}~\cite{Bagheri-2018-ID5240}.
This can be seen more clearly in Figure~\ref{fig:streamer length}, which compares the streamer length predicted by the FEDM code with the one obtained by other groups as a function of time.
Note that the streamer length is determined here as the difference between the initial seed position of the streamer and the respective point of the maximum field on the symmetry axis.

From this benchmark study, it can be concluded that the FEDM code is capable of modelling challenging discharge phenomena with high reliability.

\subsection{Abnormal glow discharge test case}
The third test case refers to an abnormal glow discharge in argon at low pressure~\cite{Becker-2009-ID2678}. It was chosen to illustrate more capabilities of the FEDM code and its further verification.
The same parallel-plate geometry as in the first case study (see Figure~\ref{fig:geometry} in Section~\ref{sec:tof-experiment}) was used with the electrode radius $R = 2$\,cm and distance $d = 1$\,cm.
Again, the problem was presumed to be axisymmetric and solved in cylindrical coordinates.
The voltage $U_\mathrm{a} = U_0 (1-\mathrm{e}^{-t/\tau})$ was applied at the powered electrode, where $U_0 = -250$\,V and $\tau = 1$\,ns.
The gas pressure was set to $p = 1$\,Torr and a constant gas temperature of $T_\mathrm{gas} = 300$\,K was assumed.

\begin{figure*}[!t]
	\centering
	\includegraphics[width=16cm]{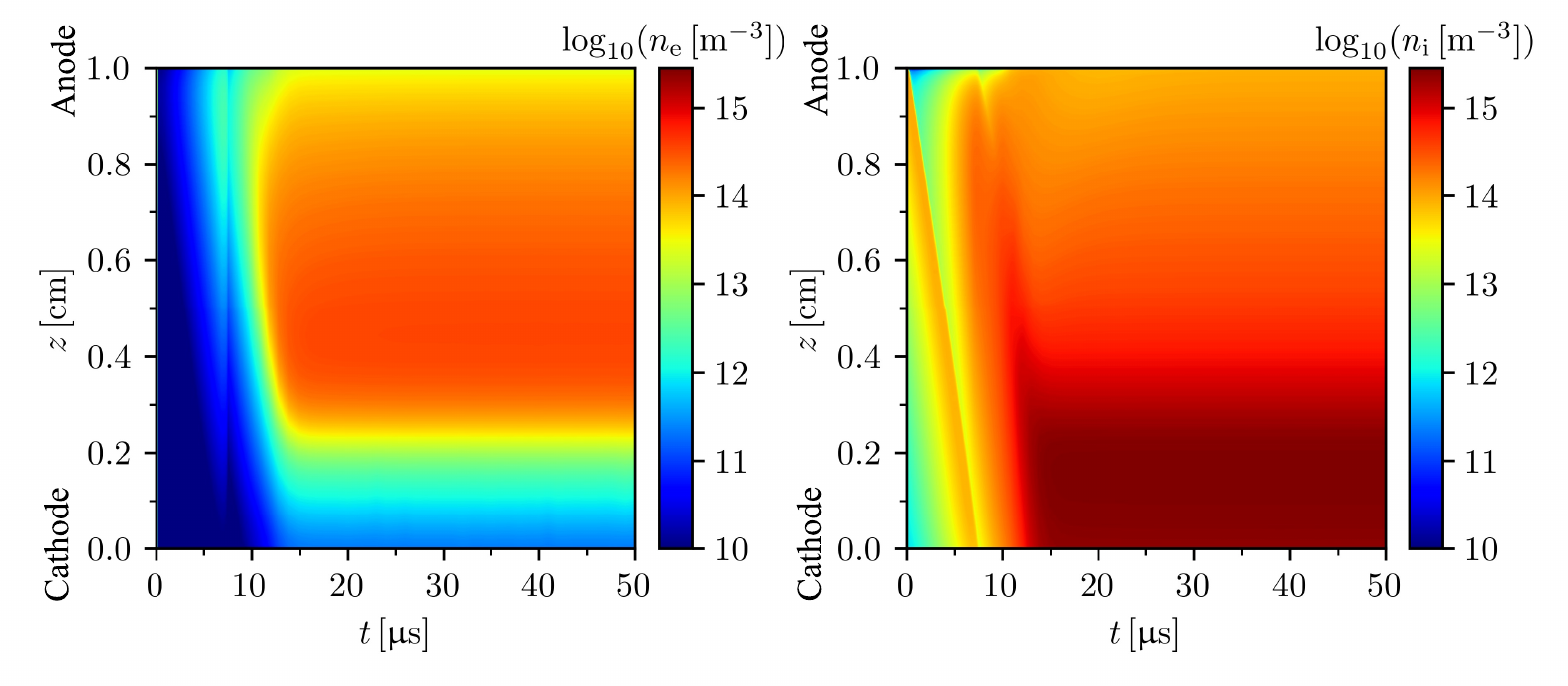}
	\caption{FEDM simulation results for the spatiotemporal evolution of electron (left) and ion (right) density at the example of an abnormal glow discharge in argon.}
	\label{fig:glow-discharge-ne-ni}
\end{figure*}

\begin{figure}[!t]
	\centering
	\includegraphics[width=8cm]{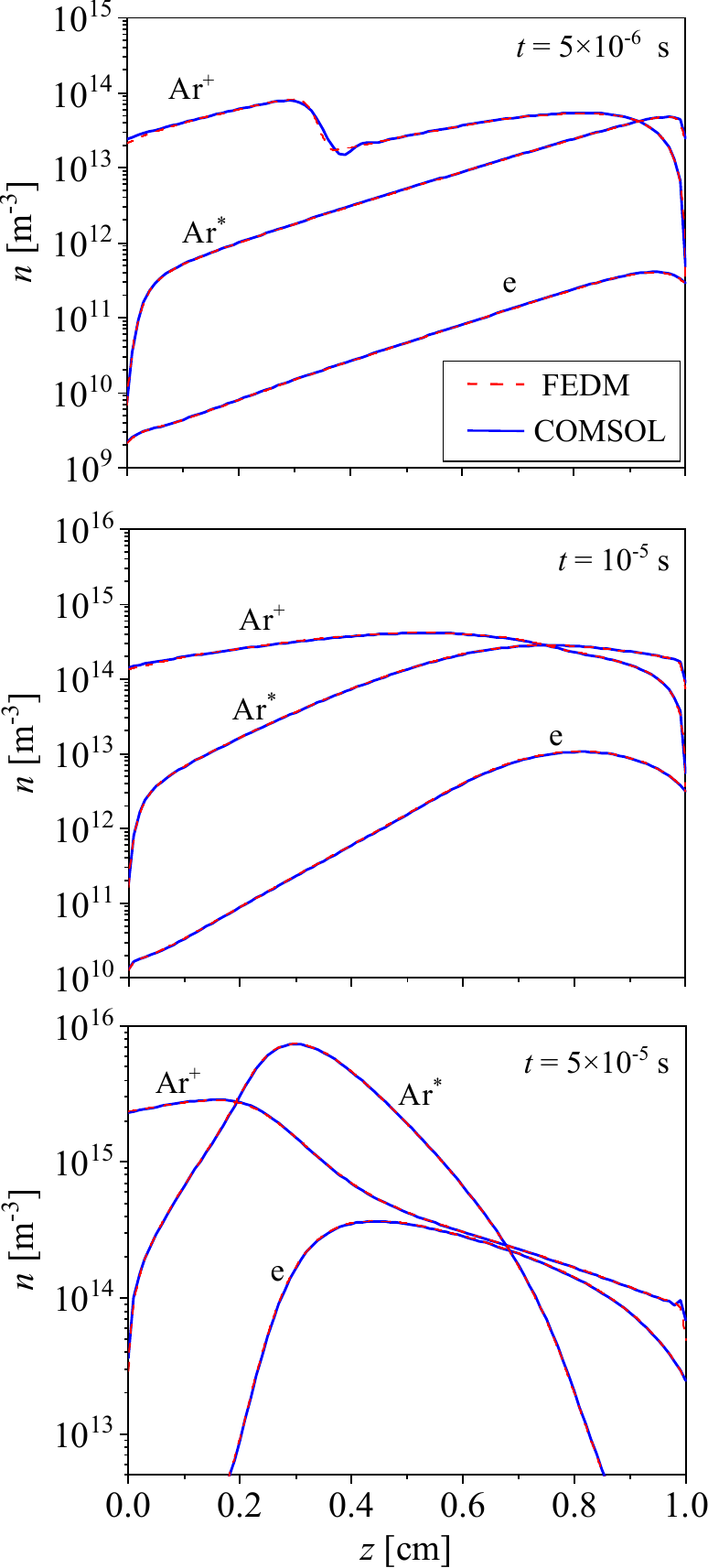}
	\caption{Axial density profiles for electrons, ions and the excited argon states at different times during ignition of the abnormal glow discharge in argon. Solid blue lines are profiles calculated using COMSOL Multiphysics\textsuperscript{\textregistered}, while dashed red lines are calculated using the FEDM code.}
	\label{fig:glow-discharge-comparison}
\end{figure}

In order to model the ignition process of the abnormal glow discharge similar to the study reported in Ref.~\cite{Becker-2009-ID2678}, a coupled solution of the whole system of equations~\eqref{eq:poisson-weak}--\eqref{eq:we-weak} with the particle fluxes~\eqref{eq:dda} and the electron energy flux~\eqref{eq:Energy flux} is required.
Here, balance equations for excited argon atoms $\mathrm{Ar}^*$, ions $\mathrm{Ar}^+$ and electrons were taken into account and the same reaction kinetic scheme as in~\cite{Becker-2009-ID2678} was used.
It includes seven processes, namely electron-impact excitation, electron-impact de-excitation, direct and stepwise electron-impact ionisation, chemoionisation, radiative de-excitation and elastic collisions.
The aforementioned automated implementation of the model was realised by means of the functionalities provided by FEDM and using the same transport and rate coefficients as in Ref.~\cite{Becker-2009-ID2678}.
It should be noted that the electron energy transport coefficients were defined as $\tilde{b}_\mathrm{e} = 5 b_\mathrm{e}/ 3$  and $\tilde{D}_\mathrm{e} = 5 D_\mathrm{e}/ 3$.

The set of boundary conditions~\eqref{eq:Particle flux for heavy particles}--\eqref{eq:Energy flux for electrons bc} was used for all species at both electrodes.
The value of the secondary electron emission coefficient was fixed at $\gamma = 0.06$, and the values of the reflection coefficients were given as $r_\mathrm{e} = r_\mathrm{exc} = 0.3$ for electrons and excited atoms, and $r_\mathrm{i} = 5 \times 10^{-4}$ for ions.
The mean energy of secondary electrons was assumed to be  $u_\mathrm{e}^\gamma = 5$\,eV.
Zero-flux boundary conditions were applied at the radial boundaries of the computational domain, effectively reducing the problem to 1D (no radial gradients).
A constant initial density of $n_0 = 10^{12}$\,m$^{-3}$ for all species (except for the constant background gas density given by $N=p/(k_\mathrm{B}T_\mathrm{gas})$) was assumed as initial condition. The initial mean electron energy was $u_\mathrm{e0} = 3 \, \mathrm{eV}$.

A structured mesh consisting of 40000 triangular elements was used, which was generated by using a built-in function of FEniCS.
As in the previous cases, linear Lagrange elements were employed for spatial discretisation. The time discretisation was realised by the variable step-size second-order BDF scheme with minimum time step $\Delta t_\mathrm{min} = 1 \, \mathrm{fs}$ and maximum time step $\Delta t_\mathrm{max} = 10 \, \mathrm{ns}$. The electron energy density was used for error control of the time stepping procedure in this case study.
The logarithmic form of the particle balance equations was used here as well.
The discretised system of partial differential equations was solved in a fully coupled manner using the same solver as for the streamer benchmark case (cf.\ section~\ref{sec:streamer-benchmark}) and relative tolerance $\mathrm{TOL} = 5 \times 10^{-4}$ as a termination criterion for the iterations during each time step. The calculations were performed on a compute node with 16 physical cores and lasted about 38 minutes. 
Following the flowchart in Figure~\ref{fig:flow_chart}, the calculations  were iterated over time until $T_\mathrm{final}=50$\,$\upmu$s. This is illustrated in Figure~\ref{fig:glow-discharge-ne-ni}, showing the spatiotemporal evolution of the charge carrier densities.
In agreement with the results presented in Ref.~\cite{Becker-2009-ID2678}, it can be seen that during the first microseconds initial seed ions drift towards the cathode, resulting in secondary electrons being emitted there by ion bombardment, which then multiply in the gap and accumulate in front of the anode.
Around $t=10 \, \upmu \mathrm{s}$, transition from the Townsend pre-phase to the abnormal glow discharge phase takes place and the stationary state is reached shortly after $t=20 \, \upmu \mathrm{s}$.

Further verification of the FEDM code is performed by comparing its simulation results with results of COMSOL Multiphysics\textsuperscript{\textregistered} simulations for the same setup.
The same input data and mesh was used in both modelling tools and automated implementation of the COMSOL model was realised by the MCPlas toolbox~\cite{Jovanovic-2021-ID5864}. Here, linear Lagrange elements were used to spatially discretise the simulation domain and the time discretisation was done using the second-order BDF method. The resulting system of nonlinear equations was solved using the constant Newton method, with MUMPS as the linear solver. The relative tolerance of the nonlinear solver was set to $10^{-4}$, and the maximum number of nonlinear iterations was limited to 40.
Figure~\ref{fig:glow-discharge-comparison} shows the axial profiles of all species densities at characteristic times during discharge ignition as obtained by use of the FEDM code and COMSOL, respectively.
Very good agreement between the results of the two simulation frameworks is obtained.

The results represented in this subsection confirm again that the FEDM code can be used for the self-consistent simulation of different phases of electric discharges with high reliability of the numerical results.

\section{Performance testing}
The performance of codes for the simulation of electric discharges is of great importance, since often large time scales have to be covered and calculations can take several days or even weeks.
To test the parallel performance of the FEDM code, the calculation time of the streamer benchmark case discussed in Section~\ref{sec:streamer-benchmark} was measured for different numbers of cores.
Based on this, the speed-up factor was determined as the ratio of the computing times measured for the use of one and several cores.
In order to gain insight into the speed-up that can be expected from commercial software packages, the same performance study was executed using COMSOL Multiphysics\textsuperscript{\textregistered}.
The first set of calculations was carried out on a server with two Intel\textsuperscript{\textregistered} Xeon\textsuperscript{\textregistered} X5570 @ 2.93\,GHz processors, with a total of 8 physical cores and 47\,GB of RAM available.
To allow comparison of the FEDM code and COMSOL, calculations were performed using similar unstructured meshes (note that the use of exactly the same mesh for both codes was not possible due to problems of interoperability).
Both meshes consisted of $500000$ elements and the same constant time step $\Delta t = 5 \, \mathrm{ps}$ was used. The Newton-based nonlinear solver was used in both cases, where direct solvers were used for each Newton iteration (Parallel Direct Sparse Solver (PARDISO) in COMSOL and MUMPS in FEDM). The global relative tolerance in COMSOL was set to $10^{-4}$ with a tolerance factor of 0.1. The relative tolerance of the nonlinear solver in FEDM was set to $10^{-4}$. The particular solvers were chosen to achieve the best overall performance. 
To reduce the overall calculation time of the performance study, the simulations were stopped at $t = 500 \, \mathrm{ps}$ in all cases.

\begin{figure}[!htbp]
\centering
\includegraphics[width=8cm]{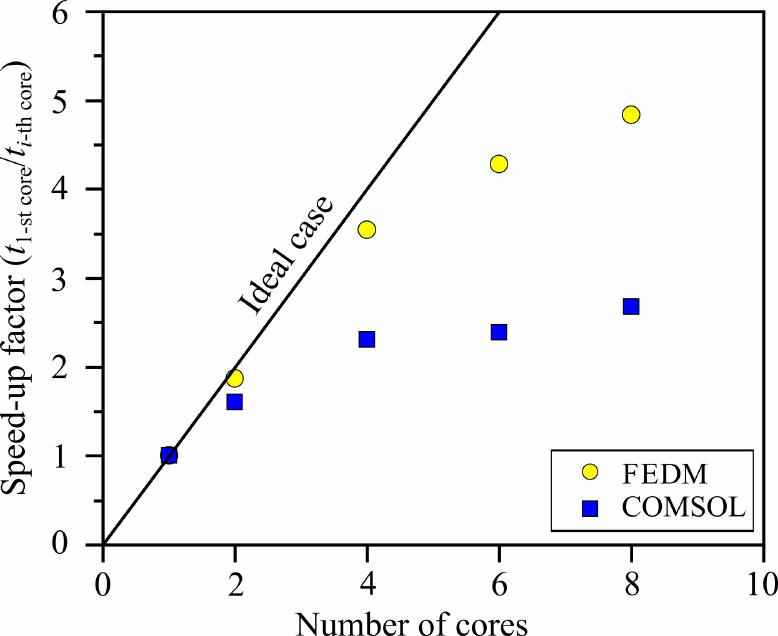}
\caption{Comparison of the speed-up factor for the FEDM code and the commercial software package COMSOL Multiphysics\textsuperscript{\textregistered} obtained for the streamer benchmark case. Calculations on a single core lasted $53 \, \mathrm{min}$ using COMSOL and  $1 \, \mathrm{h} \, 23 \, \mathrm{min}$ using FEDM.}
\label{fig:performanceFvC}
\end{figure}

Figure \ref{fig:performanceFvC} compares the speed-up factors obtained by the FEDM code and COMSOL.
FEDM clearly outperforms the latter on the given test problem, which could be due to the different ways in which the two tools are parallelised.
Therefore, the results for the FEDM code are very good and indicate that it can be executed efficiently, especially on a large number of cores. However, it is fair to mention that for single-core calculations, the calculation time for COMSOL was shorter than for the FEDM code ($53 \, \mathrm{min}$ for COMSOL and $1 \, \mathrm{h} \, 23 \, \mathrm{min}$ for the FEDM code).

\begin{figure}[!htbp]
\centering
\includegraphics[width=8cm]{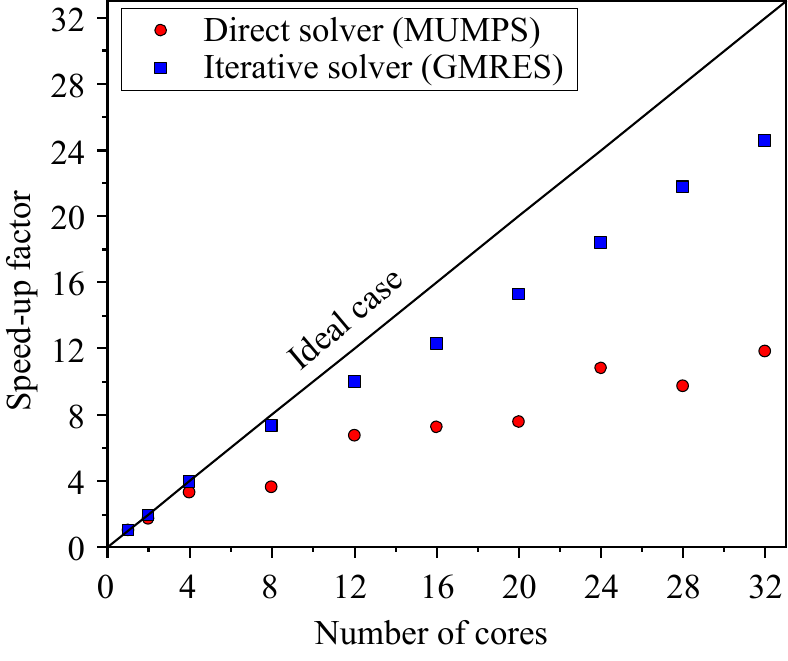}
\caption{Performance test of the FEDM code obtained on two nodes with 32 physical cores, comparing different linear solvers.}
\label{fig:performance_clusterN1}
\end{figure}

To further test the performance of the FEDM code, the simulations were repeated on a high-performance computer cluster.
Each node had a dual-socket motherboard with 8-core Intel\textsuperscript{\textregistered} Xeon\textsuperscript{\textregistered} Gold 5217 @ 3\,GHz CPUs per socket, and in total 96\,GB of RAM per node.
Nodes were connected using Mellanox\textsuperscript{\textregistered} InfiniBand MCX555A-ECAT ConnectX\textsuperscript{\textregistered}-5 VPI Adapter Cards over a Mellanox\textsuperscript{\textregistered} MSB7890-ES2F switch.
For this performance test, two nodes with in total 32 physical cores and 192\,GB of RAM were used.
The operating system was Xubuntu 20.04, and FEniCS 2019.1.0 was installed on both nodes.
The performance test was repeated as described above. Here, the speed-up was also compared when using a direct and an iterative linear solver, respectively.
In both cases, the nonlinear solver was set up to be the Newton based nonlinear solver from PETSc SNES.
For the direct solver, MUMPS was used. For the iterative solver, GMRES (generalised minimal residual)  was used with hypre AMG \cite{hypre-web-page} as preconditioner.
The speed-up factors obtained when using the direct and the iterative solver, respectively, are presented in Figure~\ref{fig:performance_clusterN1}.
For the direct solver, the speed-up factor reaches approx.\ 12 for 32 cores, while for the iterative solver a significantly higher speed-up of approx.\ 25 is achieved for 32 cores.
From this it can be concluded that the use of the FEDM code with an iterative linear solver offers great potential when used on cluster systems.

\section{Summary}
The newly developed FEDM code for the modelling of electric discharges was introduced and verified using three different case studies.
The code utilises the finite element method for solving Poisson's equation coupled with the balance equations for an arbitrary number of particle number densities of the species and the mean electron energy (where applicable) using the open-source computing platform FEniCS. The Python-based interface of FEniCS allows a smooth start to discharge modelling without the need for prior knowledge of low-level programming languages, which is sometimes required in the case of other software. It also allows better interoperability with data science software widely available in Python programming language, making the post-processing of the results easier. Furthermore, the FEDM code offers the user the possibility to define the input parameters and run the simulation without adjusting the solver. In combination with the built-in FEniCS functions, it also provides the ability to access low-level objects required for full control over the solution procedure.  For example, changing the element choice, adding numerical stabilisation, or adjusting the solver parameters for better convergence. Additional functionality, such as gas flow modelling, can be added using existing FEniCS libraries.

It was described in how far the FEDM code extends the functionality of FEniCS by adding an automated implementation of fluid-Poisson models in variational form for an arbitrary number of species.
Furthermore, a variable step-size BDF time discretisation scheme was implemented on top of FEniCS to overcome the problem of stiff equations occurring when complex plasma chemistry models have to be taken into account.
An evolutionary PID time-stepping controller was employed to implement the time-step adaptivity.

The FEDM code is structured in a way to be easily adaptable for different conditions and aspects of electric discharges. This was illustrated using three test cases, namely a time-of-flight experiment, a positive streamer in atmospheric-pressure air and a low-pressure abnormal glow discharge in argon.
Based on these case studies, the code was verified using the method of exact solutions, as well as benchmarking.
The former allowed a rigorous determination of the mesh and time order-of-accuracy.
The results showed good agreement with the theoretical values, thus verifying the code.
Benchmarking was used as a consistency check for coupled equations and also achieved a good agreement with the reference data, thus verifying the full model implementation and the code itself.

Modelling and simulation of the abnormal glow discharge in argon at low pressure was performed to show that the FEDM code can tackle various discharge configurations including different phases of electric breakdown.
Direct comparison of simulation results with data obtained by use of the commercial software package COMSOL Multiphysics\textsuperscript{\textregistered} showed very good agreement, further verifying the newly developed code.

Furthermore, a performance study was conducted.
The comparison of speed-up factors for the FEDM code and COMSOL pointed out that the FEDM code performs very well and is capable of benefiting from a high number of cores on high-performance computing clusters, especially if an iterative linear solver is used.

\section*{Acknowledgments}
This work was funded by the Deutsche Forschungsgemeinschaft (DFG, German Research Foundation)---project number 407462159. The authors wish to thank Dr.\ Peter Hill and Dr.\ Liam Pattinson of the PlasmaFAIR project for carrying out the health check and proposing and implementing improvements to the code. This support of PlasmaFAIR, funded by EPSRC (grant no.\ EP/V051822/1), is gratefully acknowledged.

\section*{Code availability statement}
The source code and input data for the models used to produce the results and analyses presented in this manuscript are publicly available at the following URL/git repository: \url{https://github.com/AleksandarJ1984/FEDM} (git commit: 5a6a617).

\section*{References}
\bibliographystyle{iopart-num}
\bibliography{mybib}

\providecommand{\newblock}{}
\begin{thebibliography}{10}
\expandafter\ifx\csname url\endcsname\relax
  \def\url#1{{\tt #1}}\fi
\expandafter\ifx\csname urlprefix\endcsname\relax\def\urlprefix{URL }\fi
\providecommand{\eprint}[2][]{\url{#2}}

\bibitem{Massines-2012-ID3544}
Massines F, Sarra-Bournet C, Fanelli F, Naudé N and Gherardi N 2012 {\em
  Plasma Process. Polym.\/} {\bf 9} 1041--1073

\bibitem{Cvelbar-2018-ID5129}
Cvelbar U, Walsh J~L, Černák M, de~Vries H~W, Reuter S, Belmonte T, Corbella
  C, Miron C, Hojnik N, Jurov A, Puliyalil H, Gorjanc M, Portal S, Laurita R,
  Colombo V, Schäfer J, Nikiforov A, Modic M, Kylian O, Polak M, Labay C,
  Canal J~M, Canal C, Gherardi M, Bazaka K, Sonar P, Ostrikov K~K, Cameron D,
  Thomas S and Weltmann {\relax K-D} 2018 {\em Plasma Process. Polym.\/} {\bf
  16} 1700228

\bibitem{Brandenburg-2018-ID5134}
Brandenburg R, Bogaerts A, Bongers W, Fridman A, Fridman G, Locke B~R, Miller
  V, Reuter S, Schiorlin M, Verreycken T and Ostrikov K~K 2018 {\em Plasma
  Process. Polym.\/} {\bf 16} 1700238

\bibitem{vonWoedtke-2013-ID3925}
von Woedtke T, Reuter S, Masur K and Weltmann {\relax K-D} 2013 {\em Phys.
  Rep.\/} {\bf 530} 291--320

\bibitem{Weltmann-2017-ID4788}
Weltmann {\relax K-D} and von Woedtke T 2017 {\em Plasma Phys. Control.
  Fusion\/} {\bf 59} 014031

\bibitem{Bekeschus-2018-ID5112}
Bekeschus S, Favia P, Robert E and von Woedtke T 2018 {\em Plasma Process.
  Polym.\/} {\bf 16} 1800033

\bibitem{Bekeschus-2020-ID5451}
Bekeschus S, Kramer A, Suffredini E, von Woedtke T and Colombo V 2020 {\em
  {IEEE} Trans. Radiat. Plasma Med. Sci.\/} {\bf 4} 391--399

\bibitem{Bisag-2020-ID5452}
Bisag A, Isabelli P, Laurita R, Bucci C, Capelli F, Dirani G, Gherardi M, Laghi
  G, Paglianti A, Sambri V and Colombo V 2020 {\em Plasma Process. Polym.\/}
  e2000154

\bibitem{Graves-1970-ID516}
Graves D~B and Jensen K~F 1970 {\em {IEEE} Trans. Plasma Sci.\/} {\bf 14}
  78--91

\bibitem{Barnes-1987-ID544}
Barnes M~S, Cotler T~J and Elta M~E 1987 {\em J. Appl. Phys.\/} {\bf 61} 81--89

\bibitem{Boeuf-1987-ID545}
Boeuf {\relax J-P} 1987 {\em Phys. Rev. A\/} {\bf 36} 2782--2792

\bibitem{Lister-1992-ID3053}
Lister G~G 1992 {\em J. Phys. D: Appl. Phys.\/} {\bf 25} 1649--1680

\bibitem{Boeuf-1995-ID3041}
Boeuf {\relax J-P} and Pitchford L~C 1995 {\em Phys. Rev. E\/} {\bf 51}
  1376--1390

\bibitem{vanDijk-2009-ID2560}
van Dijk J, Kroesen G~M~W and Bogaerts A 2009 {\em J. Phys. D: Appl. Phys.\/}
  {\bf 42} 190301

\bibitem{Lowke-2013-ID3112}
Lowke J~J 2013 {\em Plasma Sources Sci. Technol.\/} {\bf 22} 023002

\bibitem{Alves-2012-ID3406}
Alves L~L and Marques L 2012 {\em Plasma Phys. Control. Fusion\/} {\bf 54}
  124012

\bibitem{Alves-2018-ID5094}
Alves L~L, Bogaerts A, Guerra V and Turner M~M 2018 {\em Plasma Sources Sci.
  Technol.\/} {\bf 27} 023002

\bibitem{Donko-2006-ID2349}
Donkó Z, Hartmann P and Kutasi K 2006 {\em Plasma Sources Sci. Technol.\/}
  {\bf 15} 178--186

\bibitem{Donko-2011-ID2668}
Donkó Z 2011 {\em Plasma Sources Sci. Technol.\/} {\bf 20} 024001

\bibitem{Petrovic-2017-ID4095}
Petrović Z~L, Marić D, Savić M, Marjanović S, Dujko S and Malović G 2017
  {\em Plasma Process. Polym.\/} {\bf 14} 1600124

\bibitem{Loffhagen-2009-ID2590}
Loffhagen D and Sigeneger F 2009 {\em Plasma Sources Sci. Technol.\/} {\bf 18}
  034006

\bibitem{Park-1990-ID656}
Park S~K and Economou D~J 1990 {\em J. Appl. Phys.\/} {\bf 68} 3904--3915

\bibitem{Hagelaar-2005-ID2276}
Hagelaar G~J~M and Pitchford L~C 2005 {\em Plasma Sources Sci. Technol.\/} {\bf
  14} 722--733

\bibitem{Grubert-2009-ID2551}
Grubert G~K, Becker M~M and Loffhagen D 2009 {\em Phys. Rev. E\/} {\bf 80}
  036405

\bibitem{Hagelaar-2000-ID1480}
Hagelaar G~J~M, de~Hoog F~J and Kroesen G~M~W 2000 {\em Phys. Rev. E\/} {\bf
  62} 1452--1454

\bibitem{Lafleur-2019-ID5334}
Lafleur T, Schulze J and Donkó Z 2019 {\em Plasma Sources Sci. Technol.\/}
  {\bf 28} 040201

\bibitem{Bonitz-2019-ID5336}
Bonitz M, Filinov A, Abraham J~W, Balzer K, Kählert H, Pehlke E, Bronold F~X,
  Pamperin M, Becker M, Loffhagen D and Fehske H 2019 {\em Front. Chem. Sci.
  Eng.\/} {\bf 13} 201--237

\bibitem{Barnes-1988-ID582}
Barnes M~S, Cotler T~J and Elta M~E 1988 {\em J. Comput. Phys.\/} {\bf 77}
  53--72

\bibitem{Becker-2013-ID3200}
Becker M~M, Hoder T, Brandenburg R and Loffhagen D 2013 {\em J. Phys. D: Appl.
  Phys.\/} {\bf 46} 355203

\bibitem{Georghiou-2005-ID2241}
Georghiou G~E, Papadakis A~P, Morrow R and Metaxas A~C 2005 {\em J. Phys. D:
  Appl. Phys.\/} {\bf 38} R303–R328

\bibitem{Sakiyama-2010-ID2625}
Sakiyama Y, Graves D~B, Jarrige J and Laroussi M 2010 {\em Appl. Phys. Lett.\/}
  {\bf 96} 1501

\bibitem{Li-2012-ID2887}
Li C, Ebert U and Hundsdorfer W 2012 {\em J. Comput. Phys.\/} {\bf 231}
  1020--1050

\bibitem{Duarte-2015-ID3541}
Duarte M, Bonaventura Z, Massot M and Bourdon A 2015 {\em J. Comput. Phys.\/}
  {\bf 289} 129--148

\bibitem{Trelles-2018-ID5275}
Trelles J~P 2018 {\em Plasma Sources Sci. Technol.\/} {\bf 27} 093001

\bibitem{permann2019moose}
Permann C~J, Gaston D~R, Andrs D, Carlsen R~W, Kong F, Lindsay A~D, Miller J~M,
  Peterson J~W, Slaughter A~E, Stogner R~H and Martineau R~C 2019 Moose:
  Enabling massively parallel multiphysics simulation (\textit{Preprint}
  \eprint{1911.04488})

\bibitem{mfem-library}
{MFEM}: Modular finite element methods library \url{mfem.org}

\bibitem{Teunissen-2018-ID5413}
Teunissen J and Ebert U 2018 {\em Comput. Phys. Commun.\/} {\bf 233} 156--166

\bibitem{FEniCS}
 2019 {FEniCS} project \url{https://fenicsproject.org}

\bibitem{FEniCS-book}
Logg A, Mardal K~A and Wells G 2012 {\em Automated Solution of Differential
  Equations by the Finite Element Method: The {FEniCS Book}\/} (Springer
  Publishing Company, Incorporated) ISBN 3642230989

\bibitem{OpenFOAM}
{OpenFOAM} \url{https://cfd.direct/openfoam/}

\bibitem{Lindsay-2016-ID5394}
Lindsay A~D, Graves D~B and Shannon S~C 2016 {\em J. Phys. D: Appl. Phys.\/}
  {\bf 49} 235204

\bibitem{Hromadka-2016-ID5393}
Hromadka J, Ibehej T and Hrach R 2016 {\em J. Phys. Conf. Ser.\/} {\bf 759}
  012066

\bibitem{Abdollahzadeh-2016-ID5410}
Abdollahzadeh M, Pascoa J and Oliveira P 2016 {\em Comput. Fluids\/} {\bf 128}
  77--90

\bibitem{Teunissen-2017-ID4242}
Teunissen J and Ebert U 2017 {\em J. Phys. D: Appl. Phys.\/} {\bf 50} 474001

\bibitem{Verma-2021-ID5677}
Verma A~K and Venkattraman A 2021 {\em Comput. Phys. Commun.\/} {\bf 263}
  107855

\bibitem{Semenov-2022-ID6018}
Semenov I and Weltmann {\relax K-D} 2022 {\em J. Comput. Phys.\/} {\bf 465}
  111378

\bibitem{Bagheri-2018-ID5240}
Bagheri B, Teunissen J, Ebert U, Becker M~M, Chen S, Ducasse O, Eichwald O,
  Loffhagen D, Luque A, Mihailova D, Plewa J~M, van Dijk J and Yousfi M 2018
  {\em Plasma Sources Sci. Technol.\/} {\bf 27} 095002

\bibitem{UFL}
Aln\ae{}s M~S, Logg A, \O{}lgaard K~B, Rognes M~E and Wells G~N 2014 {\em ACM
  Trans. Math. Softw.\/} {\bf 40} ISSN 0098-3500

\bibitem{petsc-web-page}
Balay S, Abhyankar S, Adams M~F, Brown J, Brune P, Buschelman K, Dalcin L,
  Dener A, Eijkhout V, Gropp W~D, Karpeyev D, Kaushik D, Knepley M~G, May D~A,
  McInnes L~C, Mills R~T, Munson T, Rupp K, Sanan P, Smith B~F, Zampini S,
  Zhang H and Zhang H 2019 {PETS}c {W}eb page
  \url{https://www.mcs.anl.gov/petsc}

\bibitem{petsc-user-ref}
Balay S, Abhyankar S, Adams M~F, Brown J, Brune P, Buschelman K, Dalcin L,
  Dener A, Eijkhout V, Gropp W~D, Karpeyev D, Kaushik D, Knepley M~G, May D~A,
  McInnes L~C, Mills R~T, Munson T, Rupp K, Sanan P, Smith B~F, Zampini S,
  Zhang H and Zhang H 2019 {PETS}c users manual Tech. Rep. ANL-95/11 - Revision
  3.12 Argonne National Laboratory
  \urlprefix\url{https://www.mcs.anl.gov/petsc}

\bibitem{petsc-efficient}
Balay S, Gropp W~D, McInnes L~C and Smith B~F 1997 Efficient management of
  parallelism in object oriented numerical software libraries {\em Modern
  Software Tools in Scientific Computing\/} ed Arge E, Bruaset A~M and
  Langtangen H~P (Birkh{\"{a}}user Press) pp 163--202

\bibitem{Salari-2000-ID3784}
Salari K and Knupp P 2000 Code verification by the method of manufactured
  solutions {Sandia Report SAND2000--1444, Sandia National Laboratories}

\bibitem{Turner-2017-ID4135}
Turner M~M 2017 {\em Plasma Process. Polym.\/} {\bf 14} 1600121

\bibitem{comsol}
{COMSOL Multiphysics\textsuperscript{\textregistered} version. 5.6. COMSOL AB,
  Stockholm, Sweden.} \url{www.comsol.com.}

\bibitem{Becker-2017-ID4159}
Becker M~M, Kählert H, Sun A, Bonitz M and Loffhagen D 2017 {\em Plasma
  Sources Sci. Technol.\/} {\bf 26} 044001

\bibitem{Baeva-2020-ID5434}
Baeva M, Loffhagen D, Becker M~M, Siewert E and Uhrlandt D 2020 {\em Contrib.
  Plasma Phys.\/}

\bibitem{Jovanovic-2021-ID5864}
Jovanovi\'c A~P, Stankov M~N, Loffhagen D and Becker M~M 2021 {\em {IEEE}
  Trans. Plasma Sci.\/} {\bf 49} 3710--3718

\bibitem{Zienkiewicz}
Zienkiewicz O and Taylor R 2000 {\em The Finite Element Method\/} vol~3
  (Oxford: Butterworth) ISBN 0750650508

\bibitem{Becker-2009-ID2678}
Becker M~M, Loffhagen D and Schmidt W 2009 {\em Comput. Phys. Commun.\/} {\bf
  180} 1230--1241

\bibitem{FEMtable}
Arnold D~N and Logg A 2014 Periodic table of the finite elements
  \url{https://www-users.cse.umn.edu/~arnold/femtable/}

\bibitem{Gnybida-2009-ID2570}
Gnybida M, Loffhagen D and Uhrlandt D 2009 {\em {IEEE} Trans. Plasma Sci.\/}
  {\bf 37} 1208--1218

\bibitem{Ponduri-2016-ID3865}
Ponduri S, Becker M~M, Welzel S, van~de Sanden M~C~M, Loffhagen D and Engeln R
  2016 {\em J. Appl. Phys.\/} {\bf 119} 093301

\bibitem{HairrerNonStiff}
Hairer E, Nørsett S~P and Wanner G 1993 {\em Solving Ordinary Differential
  Equations I\/} (Berlin: Springer Verlag) ISBN 978-3-642-08158-3

\bibitem{Celaya-2014-ID5395}
Alberdi~Celaya E, Aguirrezabala J~J~A and Chatzipantelidis P 2014 {\em Procedia
  Comput. Sci.\/} {\bf 29} 1014--1026

\bibitem{Zhuang-1994-ID5401}
Zhuang M and Mathis W 1994 Research on stepsize control in the {BDF} method for
  solving differential-algebraic equations {\em Proceedings of {IEEE}
  International Symposium on Circuits and Systems - {ISCAS} '94\/} vol~5 pp
  229--232

\bibitem{tsm_course}
Moeller M 2015 {Time stepping methods, ATHENS course: Introduction into Finite
  Elements} {Delft Institute of Applied Mathematics, TU Delft}

\bibitem{Welland2014}
Welland M~J, Wolf D and Guyer J~E 2014 {\em Phys. Rev. E\/} {\bf 89} 012409

\bibitem{LIN20121225}
Lin K~M, Hung C~T, Hwang F~N, Smith M, Yang Y~W and Wu J~S 2012 {\em Comput.
  Phys. Commun.\/} {\bf 183} 1225--1236

\bibitem{Ventzek-1994-ID5407}
Ventzek P~L~G, Hoekstra R~J and Kushner M~J 1994 {\em J. Vac. Sci. Technol.
  B\/} {\bf 12} 461--477

\bibitem{HAGELAAR20001}
Hagelaar G~J~M and Kroesen G~M~W 2000 {\em J. Comput. Phys.\/} {\bf 159} 1--12

\bibitem{mumps-web-page}
 2019 {MUMPS} \url{http://mumps.enseeiht.fr/index.php?page=home}
  \urlprefix\url{http://mumps.enseeiht.fr/index.php?page=home}

\bibitem{Jovanovic-FEniCS2021}
Jovanovi\'c A~P, Loffhagen D and Becker M~M 2021 Plasma modelling using
  {FEniCS} and {FEDM} {FEniCS} 2021 conference

\bibitem{Daversin-Catty-2021-ID6140}
Daversin-Catty C, Richardson C~N, Ellingsrud A~J and Rognes M~E 2021 {\em {ACM}
  Trans. Math. Softw.\/} {\bf 47} 1--36

\bibitem{Bourdon-2007-ID3277}
Bourdon A, Pasko V~P, Liu N~Y, Célestin S, Ségur P and Marode E 2007 {\em
  Plasma Sources Sci. Technol.\/} {\bf 16} 656--678

\bibitem{Mortensen-2015-ID6141}
Mortensen M and Valen-Sendstad K 2015 {\em Comput. Phys. Commun.\/} {\bf 188}
  177--188

\bibitem{RaizerGDP}
Raizer {\relax Yu} 1991 {\em {Gas Discharge Physics}\/} vol~1 (Berlin:
  Springer) ISBN 9783642647604

\bibitem{Blevin-1984-ID2939}
Blevin H~A and Fletcher J 1984 {\em Aust. J. Phys.\/} {\bf 37} 593--600

\bibitem{Hoft-2016-ID5928}
Höft H, Becker M~M, Loffhagen D and Kettlitz M 2016 {\em Plasma Sources Sci.
  Technol.\/} {\bf 25} 064002

\bibitem{Geuzaine-2009-ID5399}
Geuzaine C and Remacle J~F 2009 {\em Int. J. Numer. Meth. Eng.\/} {\bf 79}
  1309--1331

\bibitem{hypre-web-page}
{\sl hypre}:\ {H}igh {P}erformance {P}reconditioners
  \url{http://www.llnl.gov/CASC/hypre/}

\end{thebibliography}

\end{document}